\documentclass{sig-alternate-10pt}

\RequirePackage{fix-cm}
\usepackage[letterpaper, margin=0.85in, top=0.87in]{geometry}

\usepackage{ucs}
\usepackage{array}
\usepackage[utf8x]{inputenc}
\usepackage{amsmath}
\usepackage[linesnumbered,ruled]{algorithm2e}
\usepackage{algorithmicx}
\usepackage{makecell}
\usepackage{graphicx}
\usepackage{multirow}
\usepackage{layout}
\usepackage{setspace}
\usepackage{listings}
\usepackage{mathtools}
\usepackage[]{algorithm2e}
\usepackage{graphicx}

\usepackage{float}

\floatstyle{ruled}
\newfloat{program}{tp}{lop}
\floatname{program}{Program}

\usepackage{todonotes}

\usepackage{enumitem}
\usepackage{url}

\makeatletter
\newcommand\notsotiny{\@setfontsize\notsotiny{6.5}{8}}
\makeatother

\usepackage{color, colortbl}

\definecolor{Gray}{gray}{0.9}
\definecolor{LightCyan}{rgb}{0.88,1,1}

\definecolor{dkgreen}{rgb}{0,0.6,0}
\definecolor{gray}{rgb}{0.5,0.5,0.5}
\definecolor{mauve}{rgb}{0.58,0,0.82}

\usepackage{pifont}

\usepackage[T1]{fontenc}
\usepackage{array}
\usepackage{makecell}
\newcolumntype{x}[1]{>{\centering\arraybackslash}p{#1}}

\usepackage{amssymb}
\usepackage{pifont}
\newcommand{\cmark}{\ding{51}}%
\newcommand{\xmark}{\ding{55}}%

\usepackage{tikz}

\usepackage{tikz}

\usepackage[T1]{fontenc}

\usepackage{rotating}
\usepackage[labelsep=space, labelfont=bf]{caption}

\usepackage{changepage}

\begin{document}


\title{Detecting and Diagnosing Energy Issues for \\Mobile Applications
}


\author{Xueliang Li$^1$ \and Yuming Yang$^1$ \and Yepang Liu$^2$  \and  John P. Gallagher$^3$ \and  Kaishun Wu$^1$ \and
	    College of Comp. Sci. and Soft. Eng., Shenzhen University$^1$ \and
	    Department of Comp. Sci. and Eng., Southern University of Science and Technology$^2$ \and
	    Department of People and Technology, Roskilde University$^3$
}


\date{Received: date / Accepted: date}
\maketitle
	
	\begin{abstract}
	
	Energy efficiency is an important criterion to judge the quality of mobile apps, but one third of our randomly sampled apps  suffer from energy issues that can quickly drain battery power. To understand these issues, we conducted an empirical study on 27 well-maintained apps such as Chrome and Firefox, whose issue tracking systems are publicly accessible. Our study revealed that the main root causes of energy issues include unnecessary workload and excessively frequent operations. Surprisingly, these issues are beyond the application of present technology on energy issue detection. We also found that 20.6\% of energy issues can only manifest themselves under specific contexts such as poor network performance, but such contexts are again neglected by present technology. 
	
	Therefore, we proposed a novel testing framework for detecting energy issues in real-world apps. 
	Our framework examines apps with well-designed input sequences and runtime contexts. To identify the root causes mentioned above, we employed a machine learning algorithm to cluster the workloads and further evaluate their necessity.
	For the issues concealed by the specific contexts, we carefully set up several execution contexts to pinpoint them. More importantly, we developed leading edge technology, e.g. pre-designing input sequences with potential energy overuse and tuning tests on-the-fly, to achieve high  efficacy in detecting energy issues.
	A large-scale evaluation shows that 91.6\% issues detected in our test were previously unknown to developers. 
	On average, these issues double the energy costs of the apps.
	Furthermore, our test achieves a low number of false positives. Finally, we show how our test reports can help developers fix the issues.

	\end{abstract}

	%

\vspace{-3mm}	
\section{Introduction} \label{section_introduction}

Recent years have seen a huge expansion of mobile devices such as smartphones and tablets. Due to the fact that mobile devices have limited battery capacity, energy efficiency has become one of the main considerations in system and app design. However,
many mobile apps on the market are not well optimized and can waste much battery power. We randomly sampled 89 open-source Android apps and found that 27 (30.3\%) of them suffer from serious software energy issues. 
Our experimental results (in Section \ref{section_experimental_evaluation}) show that there still exist a large number of hidden energy issues in the apps which were unknown to developers and which cause significant battery drain.

\begin{table}
	\scriptsize
	\centering
	\caption{\textbf{Comparison with present technology.}\label{table:compare_with_existing_work}}
	\renewcommand{\arraystretch}{1.2}
	\begin{tabular}{|r|c|c|}
		\hline
		\textbf{\centering Root Cause\qquad\qquad} & \textbf{{\tiny Present}} & {\tiny \textbf{This\,work}} \\\hline 
		Unnecessary Workload\qquad\;\;({\tiny 42.3\%})              & \xmark  &  \cmark\\
		{\tiny Excessively Frequent Operations}\,({\tiny 19.7\%})     & \xmark  &  \cmark \\
		{\tiny Wasted Background Processing}\;\;\,({\tiny 18.3\%})             & \cmark &  \cmark \\
		No Sleep\qquad\qquad\qquad\quad\;\;\;\;\;\,({\tiny 25.4\%})             & \cmark  &  \cmark   \\\hline
		\textbf{Manifestation Type\qquad} \qquad& \textbf{{\tiny Present}}  &  {\tiny \textbf{This\,work}}  \\\hline
		Simple Inputs\qquad\qquad\qquad\;\;({\tiny 6.3\%}) & \cmark  & \cmark \\
		Special Inputs\qquad\qquad\qquad({\tiny 73.0\%}) & \cmark  & \cmark \\
		Special Context\qquad\qquad\quad\;({\tiny 20.6\%}) & \xmark  & \cmark\\\hline
	\end{tabular}
	\vspace{-5mm}
\end{table}
In this paper, we aim to design an effective issue-detection technology to uncover these serious issues.
Essentially, knowing what are the root causes and manifestation of energy issues determines the design and implementation of the technology. For example, if a large number of energy issues are caused by runtime exceptions, then the issue-detection technology can be conveniently integrated to the current exception-handling framework. This motivates us to conduct an empirical study on the 27 energy-inefficient Android apps to learn the real root causes of energy issues and their manifestation in practice. Specifically, we aim to answer two research questions:

\vspace{-1mm}
\begin{itemize}[itemsep=4pt, topsep=4pt]
	\item \textbf{RQ1 (Issue Causes): }\textit{What are the common root causes of energy issues?}
	\vspace{-2mm}
	\item \textbf{RQ2 (Issue Manifestation): }\textit{How do energy issues manifest themselves in practice? } 
\end{itemize}	
\vspace{-1mm}

We studied 134 energy issues from the 27 open-source projects. Two findings inspired us to design a novel testing framework for detecting energy issues:

\vspace{-1mm}
\begin{itemize}[itemsep = 4pt, topsep = 4pt]
	\item \textbf{Finding 1 (w.r.t. \textbf{RQ1}):} we identified the main root causes of energy issues as unnecessary workload, excessively frequent operations, wasted background processing and no-sleep.
	
	\vspace{-2mm}
	\item \textbf{Finding 2 (w.r.t. \textbf{RQ2}):} we identified the main types of issue manifestation as simple inputs, special inputs and special contexts.
	
	
\end{itemize}
\vspace{-1mm}



Table \ref{table:compare_with_existing_work} compares present technology \cite{Banerjee_energy_bug_detection} and our work in terms of ability to deal with root causes and manifestations (and their proportions in the real-world).
\cite{Banerjee_energy_bug_detection} is based on an assumption that the main cause of energy issues is high ``E/U" ratio (the ratio of energy-consumption to hardware-utilization). If the ratio is high, it means energy consumption is high, while hardware utilization is low, implying that the app is energy-inefficient and suffering from energy issues. This point of view seems reasonable; the resulting technology is able to detect issues caused by wasted background processing and no-sleep, where E/U ratio is remarkably high. Our large-scale empirical study shows that the main root causes also include unnecessary workload and excessively frequent operations. These issues are beyond the application of E/U ratio since the ratio can not indicate the \textit{necessity of app workload}. Thus, these issues usually enlarge E and U simultaneously, so they do not increase the ratio. Hence, their technology loses efficacy in detecting such energy issues. 

Regarding Finding 2, present technology is capable of handling simple and special inputs. 
However, researchers have never considered special contexts (such as poor network performance) that hide as many 
as 20.6\% of energy issues. Thus the energy-saving potential of considering such factors has not been researched sufficiently.

Based on these observations, we propose a novel testing framework for effectively detecting energy issues. Generally, our framework examines apps with a large variety of well-designed input sequences and runtime contexts, which typically provoke energy issues to manifest themselves, as shown in Finding 2. 
To identify issues caused by unnecessary workload and excessively frequent operations, we employ a machine learning algorithm to classify the workloads, and further assess their \textit{necessity} according to several key criteria, such as lengths of continuous-high-power periods. If the workload is assessed \textit{unnecessary} and \textit{severely} energy consuming, then it will be identified as an energy issue. On the other hand, with respect to issues concealed by special contexts, we carefully devise several targeting runtime contexts to reveal them. 

To enhance the issue-detection efficacy of our framework, we also designed and implemented features such as scanning and analysing the source code of apps to extract the input sequences that are most likely to provoke energy issues.
The framework then explores the extracted input sequences with carefully-designed runtime contexts, which are dynamically adjusted to increase the chances of encountering energy issues.

The experimental results show that our tests can uncover a large number of serious energy issues in high-quality test apps, 91.6\% of which have never been discovered before. On average, these issues double the energy cost of the test apps. Manual verification also shows that our tests achieve a low number of false positives. 
Finally, we demonstrate how our test reports can facilitate developers in fixing the issues.

The key contributions of this paper are as followed:

\vspace{-2mm}
\begin{itemize}[itemsep = 4pt, topsep = 4pt]
	\item To the best of our knowledge, we conducted the largest-scale  empirical study on mobile-app energy issues (\textit{largest previous study \cite{Liu_empirical_perf}: 8 app subjects, 10 energy issues; this paper: 27 app subjects, 134 energy issues}). 
	Our findings can greatly benefit the research on energy issue detection and diagnosis.
	\vspace{-2mm}
	\item Inspired by the findings, we completely implemented an automated testing framework for detecting energy issues. The key technology includes extracting battery-hungry input sequences from source code, steering the test direction on-the-fly for high detection efficacy, etc. The evaluation shows that our framework can detect a significant number of unreported energy issues (76 unreported issues from 89 apps).  In contrast, present technology~\cite{Banerjee_energy_bug_detection}  detected only 10 unreported energy issues from 30 apps.  
	\vspace{-2mm}
	\item As far as we are aware, our evaluation on a testing framework for energy issue detection is of the largest scale (\textit{largest previous study \cite{Banerjee_energy_bug_detection}: 30 app subjects; this paper: 89 app subjects}). Our app subjects are also  of higher quality than previous work. Our subjects are selected considering metrics such as high popularity and  maintenence quality. However, most subjects in previous work can not meet this standard.
	Our evaluation both validates the efficacy of our framework and justifies our empirical findings.  

	
\end{itemize}
\vspace{-1mm}

In the remainder of this paper,  we first introduce the data source for empirical study in Section \ref{Section_datasource}, and discuss our findings in Section \ref{Section_EmpiricalStudy}. Inspired by the findings, we present a testing framework for detecting energy issues. The overview of the framework will be introduced in Section \ref{sec:framework}. We then talk about the detailed technology in Section \ref{section_detailed_technology}. Finally, we show the experimental setup and results in Section \ref{section_experimental_evaluation}.

\section{Data Source}\label{Section_datasource}

Open-source projects typically have publicly accessible issue tracking systems and code repositories. In the issue tracking systems, developers can post an issue report, which contains a title and a main body part, to report the symptoms of their observed bug/issue and the steps to reproduce the issue (optional).\footnote{\scriptsize We may use the terms bugs and issues interchangeably in this paper.} Following that, developers can discuss the issue and comment on the report. Those developers who are assigned to fix the issue can propose potential code revisions. Typically, after code review by other project members and further changes, such revisions will be committed to the project's code repository.

\begin{figure}
	\centering
	\includegraphics[width = 0.3\textwidth]{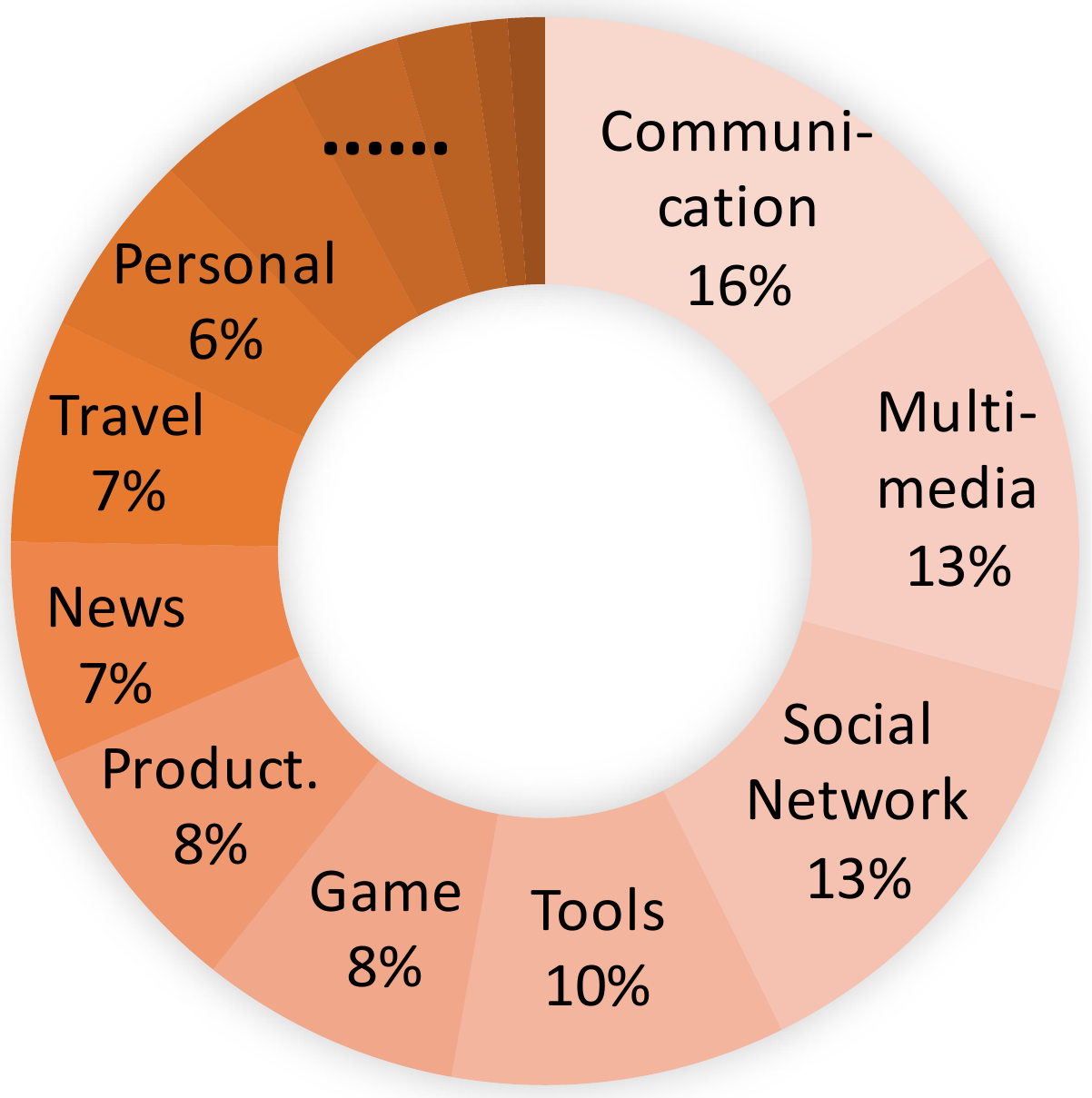}
	\caption{The 89 app subjects of  different categories in our empirical study.} \label{figure:app_candidates}
	\vspace{-6mm}
\end{figure}

Our empirical study is conducted on well-maintained Android application projects from three popular open-source software hosting platforms: GitHub\footnote{\tiny https://github.com}, Mozilla\footnote{\tiny https://dxr.mozilla.org}, and Chromium\footnote{\tiny https://www.chromium.org} repositories. The criteria for selecting app subjects for our study are: 1) a subject should have achieved at least 1,000 downloads on the market (popularity), 2) it should have more than one hundred code revisions (maintainability). Following these criteria, we randomly selected 89 app subjects from those three software hosting platforms. These open-source applications are also indexed by the F-Droid database\footnote{\tiny https://f-droid.org/}. 
Figure~\ref{figure:app_candidates} gives basic information of the 89 applications. They cover most application categories in F-Droid and in total contain 63,573 issue reports, indicating that these subjects are quite large-scale.

To search for energy issues, we employed keyword searching in issue reports' title and body to locate potential energy issues. The keywords we used are \textit{energy}, 
\textit{power} and \textit{battery}.
While keyword searching generally helps to retrieve most energy-related issue reports, it can also produce false positive results when the issue reports accidentally contain any of our keywords. To filter out such irrelevant issue reports, we manually verified each returned issue report to make sure the issue concerned is indeed an energy issue. In total, we checked 286 retrieved issue reports and this helped us locate 134 real energy issue reports from 27 apps. These energy issues were reported or updated from Jul 31st, 2009 to Jan 25th, 2019, and more than 70\% of them were reported after Jan 1st, 2014, which guarantees the timeliness of our study.

\vspace{-2mm}
\section{ Empirical Study}\label{Section_EmpiricalStudy}

To answer our research questions, we carefully studied the 134 energy issue reports. The results are as follows. 

\subsection{\textbf{RQ1:} What are the common root causes of energy issues?}\label{ssec:RQ1}

Among the 134 reports, 71 explicitly show the information on root causes of the issues. We examined all of them and observed the following six root causes. 
Some issues were caused by multiple reasons, hence, the sum of the percentages below is over 100\%.

\textbf{Unnecessary workload} ($30/71$=$42.3\%$). Many applications perform certain computations that do not deliver perceptible benefits to users. These computations incur unnecessary workload on hardware components including CPU, GPU, GPS, data storage, network interface, and screen display. For example, in the report of Chrome issue 541612\footnote{\scriptsize``Chrome" is the app name, ``541612" is the issue's ID given by the corresponding issue tracking system. }, the application produces frames constantly even when visually nothing is changed or repainted, which causes huge workload on CPU and GPU and makes rendering certain pages twice as energy-consuming as average. The report of Kontalk issue 304 shows that the app keeps pinging an address when device is disconnected from a WiFi network, continuously draining the battery. 



\textbf{Excessively frequent operations} ($14/71$=$19.7\%$). Performing certain operations too frequently can also waste battery power. In comparison with unnecessary workload, when fixing energy inefficiencies caused by excessively frequent operations, the developers do not completely remove the operations (because their functionality is necessary), but reduce the frequency of the operations. For example, in Firefox (issue 979121), whenever users type in the URL bar and the text changes, the application will  query the database (e.g. for auto-completion). Considering that users often visit the websites they visited before, developers suggested to store users' web browsing history in memory to reduce database hits to save energy. 

\textbf{Wasted background processing} ($13/71$=$18.3\%$). 
As battery-powered mobile devices are extremely sensitive to energy dissipation, it is good practice to make backgrounded applications as quiet as possible. Specifically, the ``backgound" here means that after the use of an application or ``activity" (a major type of application component that represents a single screen with a user interface\footnote{\tiny https://developer.android.com/guide/components/fundamentals.html}), users press the \textit{Home} button or switch to another application or activity, so the previous application or activity goes to background.  A typical example is Firefox issue 1022569: when users select a new tab (each tab is an activity), the invisible old tab would still keep being reloaded by itself, which wastes battery power. Later, to fix this issue, Firefox developers proposed to remove the reloading process. 


\textbf{No-sleep} ($18/71$=$25.4\%$). In contrast to wasted background processing issues, the \textit{no-sleep} issue means that when the screen is off and device is supposed to enter sleep mode, certain apps still keep the device awake. For example, Kontalk issue 143 unnecessarily holds a wake lock~\cite{Liu_wakelock}, preventing the device from falling asleep.  For another example, in Firefox (issue 1026669), the Simple Service Discovery Protocol (SSDP) is fired by the timer every two minutes when the screen is off, which not only incurs a large amount of workload but also prevents the device from entering the sleep mode. Program~\ref{program_patch_ff_1026669} gives the JavaScript patch for fixing this issue. It added the cases to deal with ``application-background" and ``application-foreground" for the SSDP service. Note that, the ``application-background" defined by developers includes both screen-off time and the scenarios where users switch to another application. So this issue belongs to two categories: \textit{no-sleep} and \textit{wasted background processing}.

\begin{program}[H]
	\begin{lstlisting}
	case "ssdp-service-found":
	- {
	-  this.serviceAdded(SimpleServiceDiscovery.findServiceForID(aData));
	-  break;
	- }
	+ this.serviceAdded(SimpleServiceDiscovery.findServiceForID(aData));
	+ break;
	case "ssdp-service-lost":
	- {
	-   this.serviceLost(SimpleServiceDiscovery.findServiceForID(aData));
	-   break;
	- }
	+ this.serviceLost(SimpleServiceDiscovery.findServiceForID(aData));
	+ break;
	+ case "application-background":
	+ // Turn off polling while in the background
	+ this._interval = SimpleServiceDiscovery.search(0);
	+ SimpleServiceDiscovery.stopSearch();
	+ break;
	+ case "application-foreground":
	+ // Turn polling on when app comes back to foreground
	+ SimpleServiceDiscovery.search(this._interval);
	+ break;
	\end{lstlisting}
	\caption{ JavaScript patch of Firefox issue 1026669.}
	\label{program_patch_ff_1026669}
\end{program}

\textbf{Spike workload} ($2/71$=$2.8\%$). A  workload spike can cause lagging UI~\cite{Liu_empirical_perf}, degrade user experience and heat up the device, inducing a huge energy waste. For instance, in RocketChat (issue 3321), when users send or receive \textit{.gif} animation pictures, CPU utilization quickly rises to 100\% and heavily affects the battery. 

\textbf{Runtime exception}  ($2/71$=$2.8\%$). In some cases, runtime exceptions may provoke abnormal behaviors of a mobile application and cause energy waste. For instance, in AntennaPod (issue 1796), the ``NullPointerException" makes the download process persist and consume power. In our study, such energy issues caused by runtime exceptions are not common and we only observed two cases.

\textit{Considering that spike workload and runtime exception are of small proportions in practice, our energy-issue detection technology only focuses on the main root causes apart from them.}

\subsection{\textbf{RQ2:} How do energy issues manifest themselves in practice?}\label{ssec:RQ2}

Out of the 134 reports, 63 contain explicit information that shows how the issues manifest themselves. We studied these 63 issues to answer RQ2. We observed that the majority of these issues require \textit{special inputs} or a \textit{special context} to trigger them, while only a few issues can be easily manifested with \textit{simple inputs}.

\textbf{Simple inputs} ($4/63$=$6.3\%$). Simple inputs mean one tap or swipe gesture in common interaction scenarios. We found four issues are of this type of manifestation.  
For example, Andlytics issue 543 lets the app refresh itself whenever the user opens the app.
And MaterialAudioBookPlayer issue 384 makes the app unnecessarily scan folders every time the user starts or leaves the app. 

\textbf{Special inputs} ($46/63$=$73.0\%$). The majority of the energy issues can only be triggered with certain specific inputs or a sequence of user interactions (e.g. text typing, taps, or swipes) under certain states of an application.  For instance, the c:geo (a geocaching app) issue 4704 requires three steps to reproduce: 1) open the app and make sure there is no GPS fix since the app starts, 2) change between cache details and other tabs of the same geocache, 3) put the device in standby and let timeout to screen-off. After a while, users would find GPS stays active even when the screen is turned off. To avoid energy waste, users decide to quit using the app.

\textbf{Special Context} ($13/63$=$20.6\%$). Special context includes environmental conditions (rather than user interactions, e.g. taps) such as the accessibility of networks, location of the device, settings of the OS and applications. In our dataset, 13 issues require such special contexts to trigger. For instance, MPDroid issue 3 appears when the user is watching stream videos but the network is disconnected; the app then keeps trying to load the video and consumes battery. AnkiDroid issue 2768 occurs when users lock the phone screen when the application is in ``review" mode and a notification comes in afterwards, so the screen will hold on until the battery is dead.



\section{Overview of Testing Framework}	\label{sec:framework}

The following observations motivated us to design an automated testing framework for effectively detecting energy issues in real apps:

\vspace{-2mm}
\begin{itemize}[topsep=4pt, itemsep=4pt]
	\item From \textbf{Finding 1}, we address previously unaddressed energy issues caused by unnecessary workload and excessively frequent operations.
	
	\vspace{-2mm}
	\item From \textbf{Finding 2}, we found that 20.6\% of energy issues can only be manifested under special context such as poor network performance. However, such factors were previously neglected.
\end{itemize}
\vspace{-2mm}

According to the first observation, as we discussed in Section \ref{section_introduction}, \textit{evaluating the necessity of app workload is crucial for identifying these issues}. 
We will use machine learning to help us cluster the workloads, and further assess its necessity, as shown later in Section \ref{section:testcase_design}.
According to the second observation, we will devise two types of most common special contexts for effectively revealing these issues, as shown in Section \ref{section_design_context}. Developers also can duplicate our experiment for detecting these hidden issues.
Importantly, we also make practical designs and implementations to enhance the efficacy of our testing framework.



\begin{figure}
	\centering
	\includegraphics[width = 0.4\textwidth]{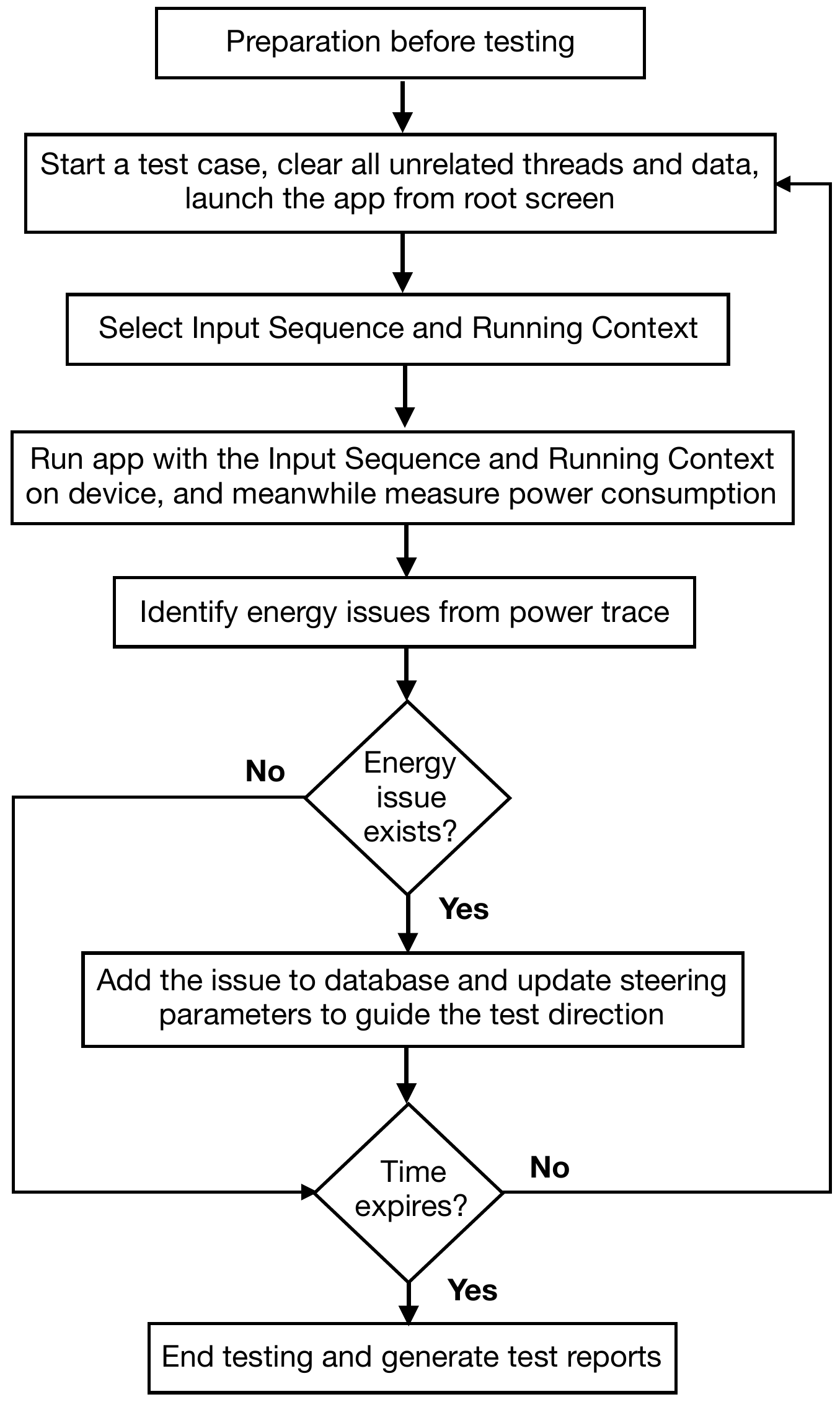}
	\caption{The flow chart for our testing framework. } \label{figure:framework}
	\vspace{-6mm}
\end{figure}

Figure \ref{figure:framework} shows the framework overview. The framework first makes sophisticated preparation before testing. For example, it inspects the source code and collects the candidate input-sequences that are most suspected of energy overuse. A set of candidate runtime contexts (containing the above mentioned two types of special contexts) are also carefully designed to increase the chance of provoking energy issues. Later, these candidate inputs and contexts will be explored under an effective and systematic scheme.


To start a test case, the framework at first clears unrelated threads and data to minimize the interference from other applications and previous test cases. Afterwards, it will select one input sequence and one runtime context from the candidates, then run the app with them. 
During the entire test, the power consumption of device is traced with a power monitor.  
Our framework will look into the power trace and decide whether an energy issue exists. If one does exist, the issue information will be added into database. The entire test is limited with a time budget. If the time budget runs out, the test will quit and test reports will be generated for developers to help fix the issues. Otherwise, our framework will start a new test case.


As mentioned above, our framework explores the inputs and contexts under a systematic scheme, where the exploring direction is tuned on-the-fly.
The rationale of our scheme is that if an energy issue occurs, it implies that the type of the input sequence and runtime context incurring this issue may be more likely to uncover energy issues than average case, since it did cause an energy issue to show up; 
we thus increase the chance of this type of inputs and context to be tested. 
Concretely, we utilize a set of parameters and iteratively update them to guide the test direction, as shown later in Section \ref{section_select_input_context}.

 The large-scale evaluation (Section \ref{section_experimental_evaluation}) shows that, exploiting these practical and targeted tests, our framework largely outperforms the state of the art on the efficacy in detecting all kinds of energy issues.
 
 \section{Detailed technology}\label{section_detailed_technology} 
 This section will introduce the detailed technology of implementing the testing framework. It involves how to design candidate input sequences and running contexts, how to steer the test direction at runtime, and how to identify energy issues from the power traces, etc. The ultimate objective of our technology is to effectively and accurately pinpoint energy issues.
 

\subsection{Preparation before testing}\label{section_preparation_before_testing}


 Our candidate input sequences are designed with high utilization of main hardware components, such as CPU, screen display and network interface, since they are usually the culprits of energy over-use, as shown in the literature \cite{Pathak_empiricalStudy_energyIssue}. At the same time, we will devise a set of artificial running contexts which are most likely to trigger energy issues.



\subsubsection{Design of candidate input sequences}\label{section_design_input}
We design two types of candidate input sequences. One is \textbf{weighted input sequences}, the other is \textbf{random input sequences}.  

\textbf{Weighted input sequences} are generated referring to the Event-Flow Graph (EFG) \cite{Event_flow_graph}. Each node in EFG is a User Interface (UI) component, such as a button or a list item. If a user interaction on a UI component, say $node_1$, can immediately activate another UI component, say $node_2$, then EFG should have a directed edge from $node_1$ to $node_2$. Technically, we utilize Layout Inspector\footnote{\tiny https://developer.android.com/studio/debug/layout-inspector} to construct the EFG of an app. So \textit{an arbitrary path in EFG is a candidate input sequence for our testing framework}. In practice, our test cases always start from the root node (i.e., the beginning UI component of the app). And the lengths of paths are constrained with a limit. Note that, Layout Inspector can construct EFG, but can not run apps with the paths. So later in testing, we use Dynodroid\footnote{\tiny https://dynodroid.github.io} to feed the paths to apps. 

More importantly, all the candidates generated from EFG are assigned with a weight.  The weight indicates the potential of a sequence to cause energy waste. And the input sequence with a larger weight has a higher priority to be tested. We adopt Equation (\ref{equation_weight}) to calculate the weight for each input sequence. $S$ is the number of system APIs (Application Program Interfaces) invoked by the input sequence. $C$ is the number of function invocations and block transitions incurred by the input sequence. $\alpha$ and $\beta$ are used for adjusting the influence of $S$ and $F$; $\alpha > 0, \beta>0, \alpha + \beta = 1$.
\begin{equation}\label{equation_weight}
weight =  \alpha * S + \beta * C
\end{equation}
\vspace{-3mm}

The reason why we resort to $S$ and $C$ to indicate the potential of causing excessive energy use is the following: as shown in \cite{Pathak_whereisenergy}, main energy-consuming components in smartphones are CPU, screen display, network interface (cellular and WiFi), further, GPS and various sensors. Except for CPU, all other components can only be controlled by system APIs. More system APIs an input sequence accesses, larger chance it may cause energy waste.  On the other hand, CPU is dedicated to executing basic operations that constitute the source code of apps, for example, arithmetic operations like additions and multiplications, and control-flow operations mainly including function invocations and block transitions. The literature \cite{source_level,source_level_practice} has shown that control-flow operations are the main energy-consumers for Java source code, which take up more than one third of the CPU energy cost. We therefore use the total number of function invocations and block transitions to indicate the potential CPU overuse of an input sequence. 

Note again that, we calculate $S$ and $F$ before testing. We first instrument the app source code, and run the app with the input sequences, and then count out their $S$ and $F$ individually.

 
Apart from \textbf{weighted input sequences}, we also designed \textbf{random input sequences} to cover some random cases we might not envisage.
We use the Monkey tool (an Android UI/application exerciser\footnote{\tiny https://developer.android.com/studio/test/monkey.html.}) to automatically generate random input sequences, such as taps and swipes. ``Random" means the position of the inputs on screen are randomly set. 
Monkey does not generate input sequences at runtime. Instead, Monkey pre-defined a large number of random input sequences. When being asked for an input sequence, Monkey will randomly deliver one of them with its ID (called \textit{seed} in Monkey). 
The advantage of this design is that testers can always use the same ID to repeat the input sequence and reproduce the test case.
 In our test, the pre-defined input sequences in Monkey are all adopted as our candidate input sequences.
 
 In summary, our test combines the two types of input sequences, namely, \textbf{weighted} and \textbf{random}. The former will help us detect energy issues in an effective means, and the latter will cover corner cases beyond our considerations. 
  Later in Section \ref{section_select_input_context}, we will show the strategy on how to balance these two types of input sequences. 
  

\subsubsection{Design of candidate running contexts}\label{section_design_context}
The entire experiment is set in a signal shielding room. We are enabled to manipulate the contextual factors, such as the strength of network and GPS signal. 
In our experiment, we designed three types of running contexts, namely, \texttt{Normal}, \texttt{Network Fail} and \texttt{Flight Mode}. In \texttt{Normal}, the network and GPS both work normally (package delivery delay is 36 ms and download bandwidth is 3.2 Mb/s). In \texttt{Network Fail}, the signal of network and GPS is seriously weak (package delivery delay lengthens to 451 ms, download bandwidth drops to 12.0 Kb/s). In \texttt{Flight Mode}, the network is closed at software level by operating system, but GPS works normally. 

The \textbf{reason} why we choose \texttt{Network Fail} and \texttt{Flight Mode} as representatives for special contexts is that, our empirical study shows they are the two major types of special contexts. The former holds 25.0\% (3 out of 12), the latter occupies 16.7\% (2 out of 12) of all issues manifested under special context.

We also designed a special type of running context, \texttt{Non-background}. We designed this context because in our experiment we observed that it can provoke more no-sleep issues, as shown later in Section \ref{section_issue_manifestation}.
In \texttt{Non-background}, the network and GPS work ordinarily, however, we do not input a press of \textit{Home} button to the device after \texttt{EXECUTION} stage (the stage-division for test cases will be explained in Section \ref{section:testcase_design}). That is, the test case does not have \texttt{BACKGROUND} stage, and straight goes to \texttt{SCREEN-OFF}.

%




 
\subsection{Steer the test direction on-the-fly}\label{section_select_input_context}

Our framework steers the test direction dynamically based on test history. Algorithm \ref{algorithm_select_input_context} shows details of our steering scheme. The rationale behind is this: when an energy issue is detected, it implies that this type of input sequence and running context may have a larger opportunity to provoke energy issues than normal case since it did bring an energy issue to light. Hence, our framework will generate slightly more of this type of test cases for larger chance of detecting energy issues. \\

\vspace{-3mm}
 \begin{algorithm}[tbh]
 	\scriptsize
 	\KwData{\\
 		$WeightedInputSequences=\{(sequence_i, weight)\}$;\\
 		$RandomInputSequences=\{sequence_j\}$;\\ $RunningContexts[N]=\{context_k\}$;\\
 		$0<p_{wei}<1$, $P_{ctx}[N]=\{0<p_k<1\}$;\\
 		$\Delta_{wei}$, $\Delta_{context}$; \\
 	}		           
 	Preparation before testing\;
 	Start a test case, clear unrelated threads and data, launch the app from root screen\;
 	\CommentSty{\#-------Select Input Sequence and Running Context-------\#}\\
 	Determine the type of input sequence, and weighted input sequence has a probability of $p_{wei}$ to be approved\;
 	\uIf{the determined type is weighted}{
 		Select an unexplored sequence with highest $weight$ in $WeightedInputSequences$\;
 	}
 	\Else{
 		Randomly select one sequence from $RandomInputSequences$\;
 	}
 	Select one context (e.g., $context_k$) from $RunningContexts$ with its corresponding probability (e.g., $p_k$)\;
 	\CommentSty{\#-----------------------------------------------------------------------------\# }\\
 	Run app with the Input Sequence and Running Context on device, and meanwhile measure power consumption\;
 	Identify energy issues from power trace\;
 	\If{ there exists an energy issue}{
 		Add the energy issue to database\;
 		\CommentSty{\#----------------Update steering parameters---------------\#}\\
 		\uIf{ the energy issue is triggered with a weighted input sequence \textbf{and} $p_{wei} <= \texttt{wei\_up\_threshold}$}{$p_{wei} := p_{wei} + \Delta_{wei}$\;}
 		\ElseIf{$p_{wei} >= \texttt{wei\_down\_threshold}$}{$p_{wei} := p_{wei} - \Delta_{wei}$\;}
 		\Switch{which context triggers the energy issue}{
 			\Case{e.g., $context_k$}{
 				\If{$p_k<= \texttt{cxt\_up\_threshold}$ \textbf{and} there are $n$ elements in $P_{ctx}$ above $\texttt{cxt\_down\_threshold}$ \textbf{and} $n>0$}{
 					$p_k := p_k + \Delta_{context}$\;
 					Decrease those n elements individually by $\Delta_{context}/n$\;}
 			}  
 		}
 		\CommentSty{\#---------------------------------------------------------------------\#}\\
 		
 	}
 	\lIf{time expires}{\textbf{End testing}, generate test reports}{Go back to \textbf{line 2}}
 	
 	\caption{Steer the test direction on-the-fly}\label{algorithm_select_input_context}
 	
 \end{algorithm}

\paragraph{\textbf{Data for the algorithm}} The candidate input sequences and running contexts are designed based on the approach we demonstrated in Section \ref{section_design_input} and \ref{section_design_context}. We present them in the data structures of {\small $WeightedInputSequences$, 
$Random$ $InputSequences$ and $RunningContexts$}. $N$ is the number of candidate running contexts. In our experiment, we devised 4 running contexts, so $N == 4$.

$p_{wei}$ and $P_{ctx}$ are the steering parameters for directing the test.
$p_{wei}$ is the probability of choosing a \textbf{weighted} input sequence for the upcoming test case. In our experiment, we initialize it as 50\%, so the test case has a chance of 50\% to run with weighted input sequence at the beginning of test, and we will update and refine $p_{wei}$ on-the-fly during the entire testing.
On the other hand, one element $p_k$  in $P_{ctx}$ represents the probability of choosing $context_k$ in {\small $RunningContexts$}. We will also update $P_{ctx}$ at runtime. Note that, the summation of the elements in $P_{ctx}$ is bound to 1.

$\Delta_{wei}$ is the increment utilized to increase or decrease $p_{wei}$ to refine $p_{wei}$. $\Delta_{context}$ plays the same role for $P_{ctx}$. Larger $\Delta_{wei}$ and $\Delta_{contex}$ we employ, more aggressively we tune the test direction. 

\vspace{-2mm}
\paragraph{\textbf{Details of the algorithm}}

We first prepare the data and initialize the parameters (i.e., $p_{wei}$, $P_{ctx}$, $\Delta_{wei}$ and $\Delta_{context}$). We then start a test case, clear unrelated threads and data, launch the app from root screen. 
Next, we will decide the type of input sequence; weighted input sequences have a probability of $p_{wei}$ to be chosen. If the chosen type is ``weighted", we then select an unexplored sequence with the highest $weight$ in {\small $WeightedInputSequences$}. Otherwise, we randomly select one from {\small $RandomInputSequences$}. Likewise for running context, we select one from {\small RunningContexts} with its corresponding probability.

We feed the app subject with the selected input sequence and running context on device, and measure the power consumption at the same time. The power trace will be analysed to confirm whether an energy issue occurs.
If there exists an energy issue, it implies that the corresponding type of input sequence and running context may be profitable for provoking more energy issues, our testing framework then steers a bit to this direction. Specifically, if it is triggered with a weighted input sequence, we increase $p_{wei}$ by $\Delta_{wei}$, and $p_{wei}$ should not exceed the $wei\_up$ $\_threshold$. Otherwise, we decrease $p_{wei}$ by $\Delta_{wei}$. Also, we keep $p_{wei} >= wei\_down$ $\_threshold$.


An analogous approach is applied to refining $P_{ctx}$. We check under which running context (e.g., $context_k$) the issue occurs, then increase its testing probability (e.g., $p_k$). However, the precondition is that there should be at least one elements (except $p_k$ itself) in $P_{ctx}$ that are above $cxt\_down$ $\_threshold$ because we, on one hand, intend to rebalance the probabilities, on the other hand, we should also let all contexts have at least a possibility of $cxt\_down\_threshold$ to be tested.

 \subsection{Identify energy issues from power trace}\label{section:testcase_design}

\begin{figure}
 	\centering
 	\includegraphics[width = 0.52\textwidth]{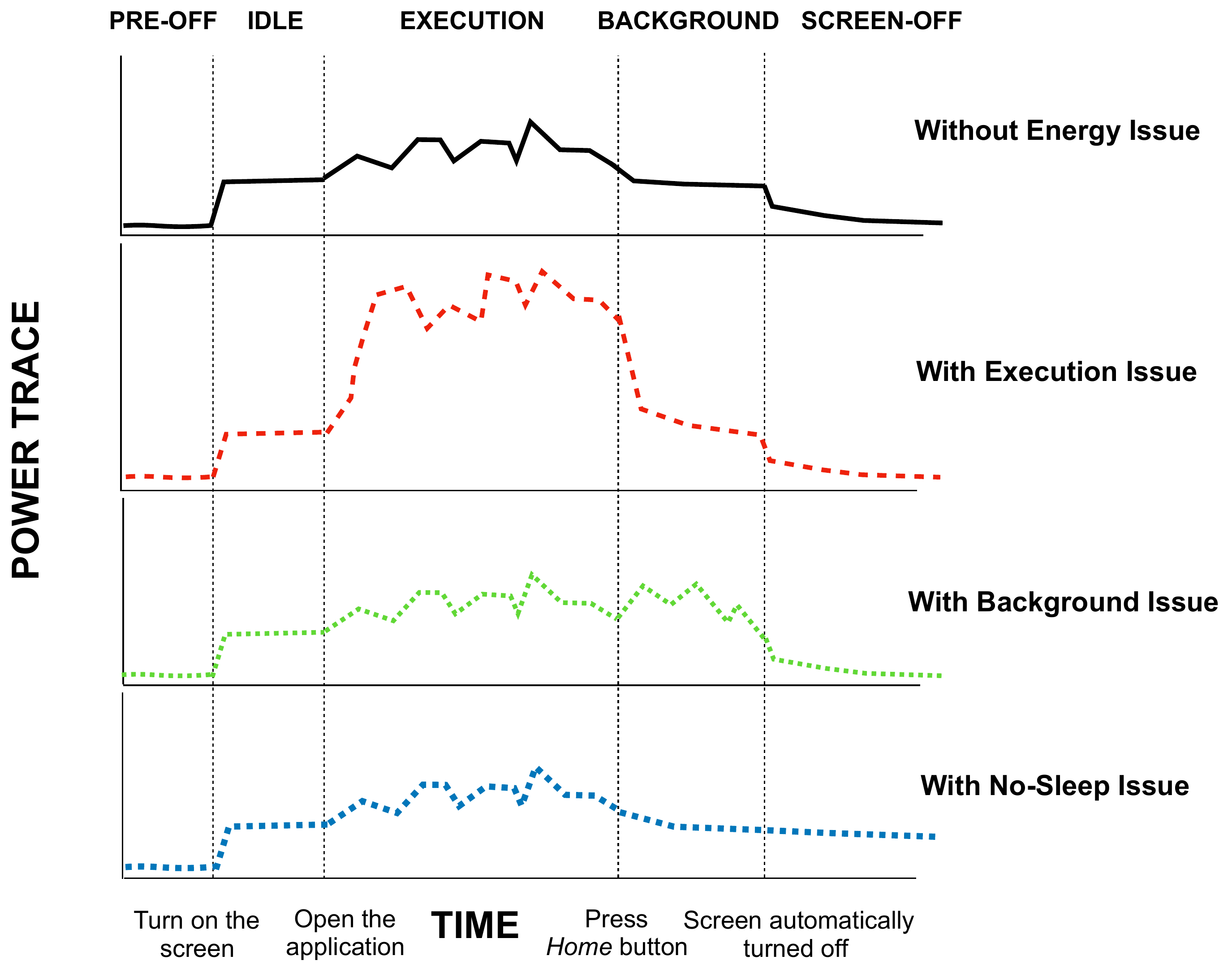}
 	\caption{An illustration of power traces with energy issues.} \label{figure:powertrace}
 \end{figure}

 We divide the power trace into five stages, namely, \texttt{PRE-OFF}, \texttt{IDLE}, \texttt{EXECUTION}, \texttt{BACKGROUND} and \texttt{SCREEN-OFF}. This division can help us identify three types of energy issues: execution issues (including issues caused by unnecessary workload and excessively frequent operations), background issues and no-sleep issues.
 Figure \ref{figure:powertrace} shows an illustration of power traces with these three types of energy issues.

 \texttt{PRE-OFF} stage is the beginning stage where the device is powered but the screen is off. Then, the test case will be transferred to \texttt{IDLE} stage by turning on the screen. To enter \texttt{EXECUTION} stage, the subject application will be opened and run with a certain input sequence and running context, which are generated as shown in Section \ref{section_preparation_before_testing} and \ref{section_select_input_context}.  After \texttt{EXECUTION} stage, the application will be fed with a press of \textit{Home} button to enter \texttt{BACKGROUND} stage. The final stage is \texttt{SCREEN-OFF} stage, which begins when the screen is supposed to be turned off automatically, however, a part of energy issues will keep the screen on even at \texttt{SCREEN-OFF} stage, eating a large amount of battery power. 
 \\

\vspace{-3mm}
\paragraph{Identifying execution issues}
For execution issues, as we discussed in Section \ref{section_introduction}, evaluating the necessity of app workload is crucial for identifying them. 
Specifically, we employ the Dbscan clustering algorithm~\cite{DbScan} (Density based spatial clustering of applications with noise) to fulfil this purpose. 
The objective of  Dbscan is to classify multidimensional data points into three groups, namely, core points, border points and outlier points. After clustering, the data points should have the following properties: 

1) For a core point, the number of its neighbours (the points within a range of $\varepsilon$ from it) is no less than a certain value, $MinPts$. Generally speaking, the core points are ``quite close and gathered".

2) For a border point, its neighbours are less than $MinPts$, but it is a neighbour of at least one core point or another border point. 

3) For an outlier point,  its neighbours are less than $MinPts$, and it does not have either a core or a border neighbour.  

We treat each test case as a data point, and treat the test cases in the same app category as a data set for clustering.
The dimensions of each data point we employed for clustering are $l_{chpp}$, $n_{chpp}$, $\mu_{chpp}$, $\mu_{exe}$, which are all extracted from power trace of \texttt{EXECUTION} stage of each test case.
$l_{chpp}$ is the total length of continuous-high-power periods. Continuous-high-power period is when power continuously exceeds a certain threshold longer than a certain length. $n_{chpp}$ is the number of these periods. $\mu_{chpp}$ is the average power of these periods. $\mu_{exe}$ is the average power of the entire \texttt{EXECUTION} stage.


Dbscan then classifies the test cases into those three groups. We label the test cases in core and border groups as ``normal", and label the ones in outlier group as suspects for suffering from energy issues.
Later, our evaluation on 89 apps (involving 35600 test cases) shows that only 1.1\% test cases are outliers, which are observably energy-consuming.
We assume they have  ``unnecessary" workloads or operations.
Finally, we manually verify whether there exist real execution issues.
In the experiment, our testing framework detected 47 candidate execution issues from those 1.1\% test cases.  Only three (out of the 47) are false positives, indicating the high reliability of this approach. In contrast,  current technology~\cite{Banerjee_energy_bug_detection} detected 3 candidate execution issues from 30 apps, and still one of them is a false positive.




\vspace{-3mm}
\paragraph{Identifying background and no-sleep issues}
 If the app is free from background issues, the power trace in \texttt{BACKGROUND} stage is supposed to be similar to that in \texttt{IDLE} stage.
We thus compute the dissimilarity value of the two traces. If the value is above a certain threshold (40\% in our experiment), we label this test case as a candidate for a background issue. 

We identify the no-sleep issues in the same way.  We compare \texttt{PRE-OFF} with \texttt{SCREEN-OFF}. If the dissimilarity exceeds a certain threshold (50\% in our experiment), we speculate this test case is suffering from a candidate  no-sleep issue.

\subsection{Manual verification}

After the candidate issues are found, we manually verify whether they are actual energy issues.
The steps for manual verification are as followed: 

 \textit{a}. We re-run the test case to check if the energy issue can be reproduced. 
 
 \textit{b}. We analyse the power trace to affirm its significant impact on energy consumption.
 
 \textit{c}. We observe symptoms of the issue, including flickering UI, popping-up messages, long CPU wake-time, high CPU utilization and high GPU rendering speed.

  \textit{d}. We inspect the execution trace (obtained from Android Traceview\footnote{\tiny https://developer.android.com/studio/profile/traceview}, now Traceview is deprecated, developers can use CPU Profiler\footnote{\tiny https://developer.android.com/studio/profile/cpu-profiler} instead) to check if there has frequently-provoked methods in the program. In practice, they are usually the suspects of defective code. According to the issue reports in our empirical study,  for the issues diagnosed using this information, the faulty code was 100\% found and fixed. Here are two examples\footnote{\tiny https://bugs.chromium.org/p/chromium/issues/detail?id=349059}$^,$\footnote{\tiny https://bugs.chromium.org/p/chromium/issues/detail?id=480522}.
  
  Only if the issue can be reproduced, cause noticeable energy waste, have visible symptoms, and its faulty code can be located, we affirm its real existence. Otherwise, we determine it as a false positive.
  
  

  \begin{table*}
  	\vspace{-3mm}
  	\footnotesize
  	\centering
  	\caption{\textbf{Examples of detected energy issues.}}\label{table:energy_issue_examples}
  	\renewcommand{\arraystretch}{1.2}
  	\begin{tabular}{|c|c|c|c|c|c|r|c|}
  		\hline
  		& &  & \textbf{Activity or}  & \textbf{Hit}  & \textbf{Context or}  &   \textbf{Energy}  & \textbf{Reported}  \\ 
  		
  		\multirow{-2}{*}{\textbf{Application}}  & \multirow{-2}{*}{\textbf{Category}} &  \multirow{-2}{*}{\textbf{Issue Type}} & \textbf{Symptom}& \textbf{Rate$^1$} &  \textbf{NonBg$^2$} & \textbf{Waste} & \textbf{before?}\\
  		\hline
  		Leisure  & News & Execution & {\scriptsize 3 \textit{.gif} playing on page} & 1.0\% & Normal & 25.9\% & No \\
  		\hline  
  		GPS Status & Travel & Execution & {\scriptsize Not obvious} & 47.0\% & Normal & 15.3\% & No \\
  		\hline
  		BatteryDog & Tools & Execution &{\scriptsize Editing lengthy text }&  1.0\% & Normal & 30.8\% & No\\
  		\hline
  		Rocket Chat & Comm. & Execution &{\scriptsize Connect to server} & 1.0\% & Normal & 12.8\% & No \\
  		\hline
  		Chrome & Browser & Execution &{\scriptsize Keep loading pages} & 4.0\% & WiFi Fail & 9.7\% & No \\
  		\hline
  		Chess Clock & Tools & Background &{\scriptsize Device heated up }& 100.0\% & WiFi Fail  & 59.2\% & No \\
  		\hline
  		Vanilla & Multimedia & No Sleep &{\scriptsize Enqueue many tracks }& 59.0\% & WiFi Fail & 242.8\% & No \\
  		\hline
  		cgeo & Travel & No Sleep &{\scriptsize App get stuck }& 2.0\% & Flight Mode & 179.4\% & Yes \\
  		\hline
  		AntennaPod & Multimedia & No Sleep &{\scriptsize Popping up messages} & 1.0\% & NonBg & 184.4\% & Yes \\
  		\hline
  		\multicolumn{8}{l}{\scriptsize 1. Hit rate here is the percentage of test cases detected having the energy issue in that app. 2. ``NonBg" is ``\texttt{Non-background}".}\\
  	\end{tabular}
  	\vspace{-2mm}		
  \end{table*}


\section{Experimental Evaluation}\label{section_experimental_evaluation} 

In this section, we first present the specifics of our experimental setup. Then, we evaluate our testing framework on various aspects, such as its efficacy of detecting energy issues, its comparison with the state-of-the-art, etc. The result shows that our testing framework largely outperforms the present technology, which benefits from the sound empirical study and targetting implementations of the testing framework.  

\subsection{Experimental setup}
\label{section_experimental_setup}


We employ the Odroid-XU4 development board\footnote{\tiny https://wiki.odroid.com/odroid-xu4/odroid-xu4}, whose processor has four big cores with a frequency of 2 GHz and four small cores with a frequency of 1.3 GHz. The main memory is as large as 2 GB. The board possesses a powerful 3D accelerator, Mali-T628 MP6 GPU. The high capacity of Odroid-XU4 board guarantees its performance for most applications on the market.

 It also provides a package of developer options. For example, it can display CPU utilization and GPU rendering profile on screen which will aid us in verifying the real existence of energy issues.
It is also equipped with a power monitor, Smartpower2\footnote{\tiny https://wiki.odroid.com/accessory/power\_supply\_battery/smartpower2}, to measure the real-time power consumption. The sampling rate is 100 Hz.
Due to these rich features, Odroid board is widely-employed in the field of energy optimization for mobile devices \cite{paper1_use_odroid}\cite{paper2_use_odroid}.


 We use Android as our target operating system since it is open-sourced and its market share captures around 85.9\%\footnote{\tiny https://www.statista.com/statistics/266136/global-market-share-held-by-smartphone-operating-systems/} of the worldwide smartphone volume by the end of first quarter of 2018.  
We evaluate our framework on the same set of 89 app subjects in the empirical study.  
Our total testing time for the 89 app subjects is 2373.3 hours, i.e., 98.9 days (on average, 1.11 days for one app subject). 





\subsection{The efficacy of our testing framework}
The experimental result shows that our test detected 91 candidate energy issues, among which we manually confirmed \textbf{83 real energy issues}. We have \textbf{8 false positives}.
The false positives are caused by the noise in power samples in some case. The noise could be a result of OS activities at background, such as Android Runtime Garbage Collection\footnote{\tiny https://source.android.com/devices/tech/dalvik/gc-debug}, random system downloading and updating.


Table \ref{table:energy_issue_examples} shows 9 examples of the detected energy issues. For instance, in \texttt{Leisure},  three animated \textit{.gif} are loaded and playing at the bottom of one certain page  even though they are invisible to users most of the time.  This execution issue wastes 25.9\% energy use. It can be fixed by freezing the animation when the \textit{.gif} pictures are not shown on the screen. For another example,  when \texttt{Chess Clock} is after use and backgrounded, the device will be heated up from $41.2^\circ$C to $60.9^\circ$C due to the inefficient and long utilization of CPU. The average power of this issue is 59.2\% higher than that of the \texttt{IDLE} stage.


 \textbf{91.6\%} (76 out of 83) detected energy issues in our test are  \textbf{newly-reported}. Our experiment also shows that these issues averagely double the energy consumption of the apps (see later in Section \ref{section_energy_waste}). Without our test, these serious energy issues would have never surfaced even though the battery drains desperately.

 
 On the other hand, 94.5\% (120 out of 127) energy issues in our empirical study were not listed in the issues detected  by our test.  
 Considering that the test is very large (2373.3 hours), it is infeasible to go through all the test cases to find the reasons why they were not detected, and present their proportions. However, 
 we conjecture the major reasons may be the following:

 Firstly, our standard of determining an energy issue is much higher than that of developers. The issues detected by our test usually have long-lasting impact on the energy consumption, the energy waste is usually above 10.0\%. However, many issues detected by developers may only cause a transient workload, and their energy waste can be hardly more than 10.0\%.
 Secondly, a number (26.7\%, 32 out of 120) of the issues are not reproducible, so our test can not trigger them either.
 Thirdly, due to the time limit, the variety of input sequences and running contexts in our test is not large enough to cover all of them.

 After all, our framework still shows its efficacy in detecting severe and unreported energy issues, which were beyond the vision of researchers formerly.

\subsection{Comparison with the state-of-the-art}
We will compare the efficacy of our testing framework with the present technology \cite{Banerjee_energy_bug_detection}. Since most of their app subjects ($18/30=60.0\%$) are not open-sourced, we cannot look into the execution trace to verify the issues.
 Also, similar to the comparison with developers, our standard of identifying energy issues is much higher than that of the former researchers, we then do not make comparison on the set of their subjects. Instead, we make it statistically.

\begin{figure}
	\centering
	\includegraphics[width = 0.3\textwidth]{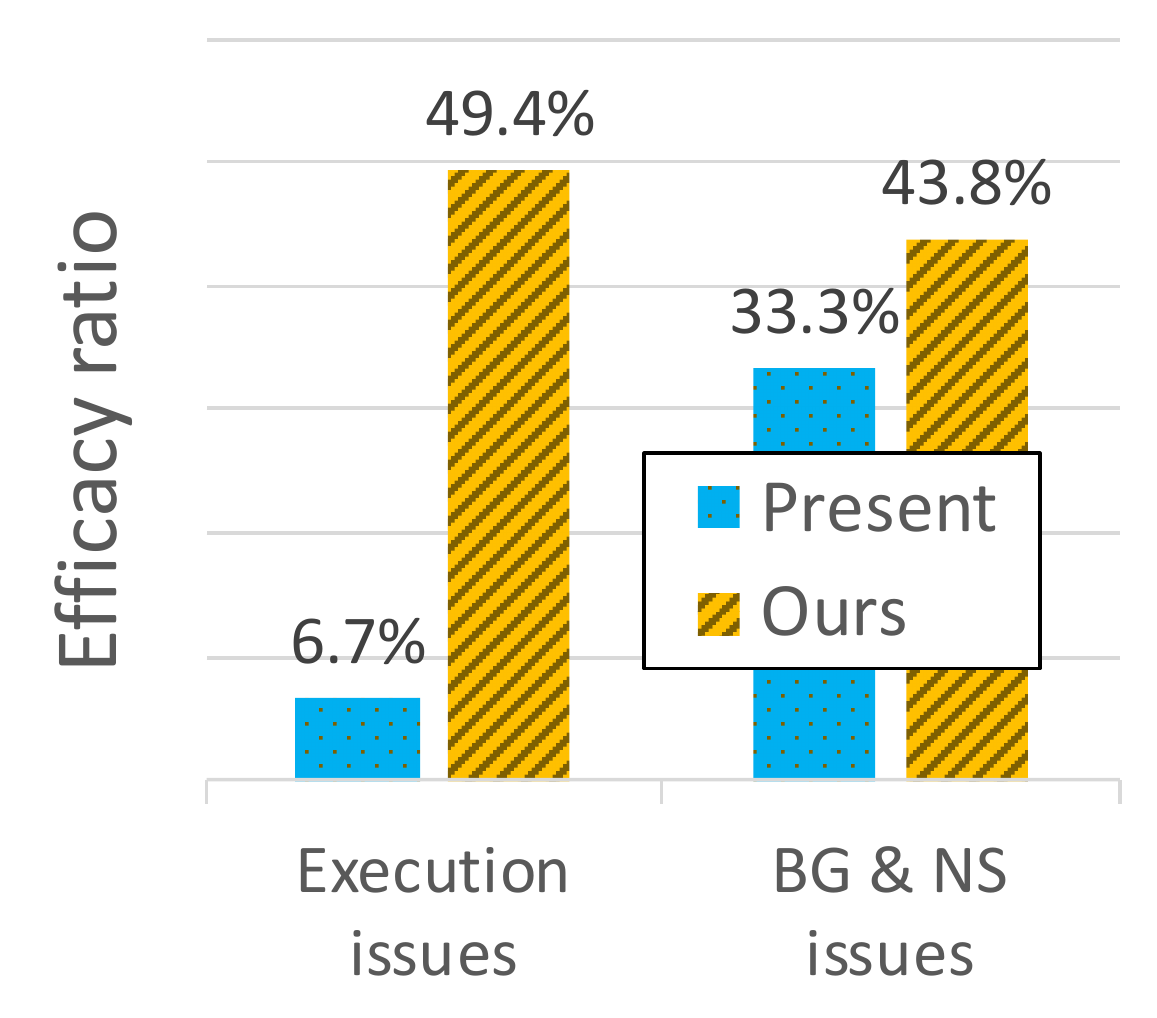}
	\caption{Comparison on issue-detection efficacy ($r_s^d$) with the present technology.} \label{figure:framework_efficacy}
\end{figure}

We use the efficacy ratio, i.e. the ratio ($r^d_s$) of the number of detected issues to the number of subject apps, to indicate the efficacy of a testing framework. As shown in Figure \ref{figure:framework_efficacy}, for execution issues, their $r^d_s$ is $6.7\%$ (2 issue out of 30 apps), our $r^d_s$ is $49.4\%$ (44 out of 89), for background (BG) and no-sleep (NS) issues together, their $r^d_s$ is $33.3\%$ (10 out of 30), ours is $43.8\%$ (39 out of 89). 
We put background and no-sleep issues together since their work did not distinguish them. 
 The comparison validates that our testing framework is more effective than the state-of-the-art technology. It is because our work is based on the insightful empirical findings, rather than an ungrounded assumption.    

\begin{figure}
	\centering
	\includegraphics[width = 0.45\textwidth]{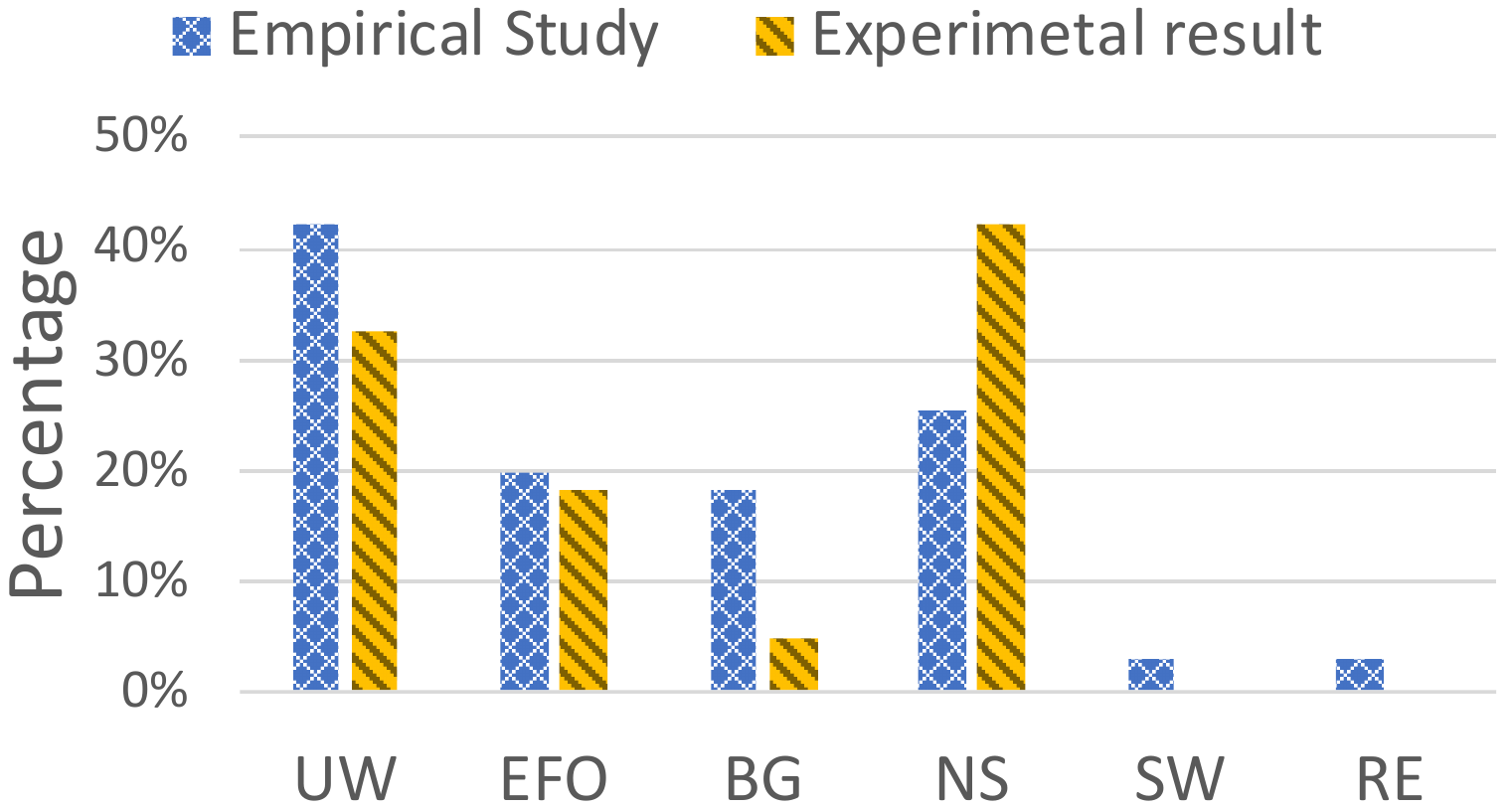}
	\caption{Breakdown of issue causes in \textit{empirical study} and \textit{experiment}.} \label{figure:exp_emp_energy_issues}
\end{figure}

\subsection{Issue cause and manifestation}\label{section_issue_manifestation}

Figure \ref{figure:exp_emp_energy_issues} demonstrates breakdown of energy issues of different causes in \textit{empirical study} and \textit{experiment}. ``UW" is unnecessary workload, ``EFO" is excessively frequent operations, ``BG" is wasted background processing, ``NS" is no-sleep, ``SW" is spike workload, ``RE" is runtime exception. 
It shows that UW and EFO are very significant root causes of energy issues in both the \textit{empirical study} and the \textit{experiment}, which justifies our Finding 1.

\begin{figure}
 	\centering
 	\includegraphics[width = 0.47\textwidth]{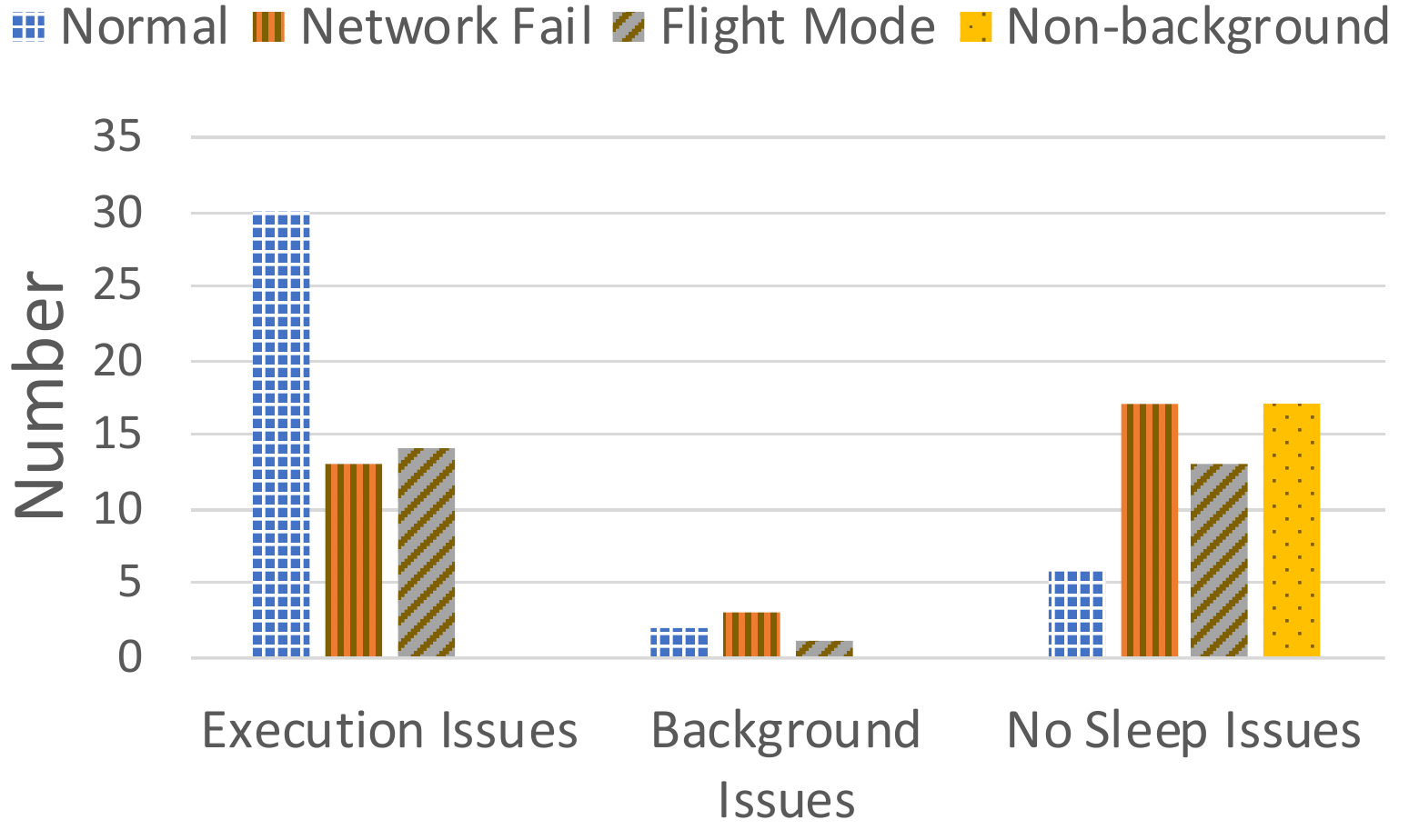}
 	\caption{Energy issues manifested under different context.} \label{figure:distribution_detected_energy_issues}
 	\vspace{-3mm}
 \end{figure}

 Figure \ref{figure:distribution_detected_energy_issues} presents the numbers of energy issues provoked under different context, which captures manifestation characteristics of energy issues. We can see that most execution issues are manifested under the \texttt{Normal} context because many scenarios where issues occur require normal network and GPS context to reach. For example, as we talked above, the issue in \texttt{Leisure} occurs only when those three \textit{.gif} pictures were downloaded and showing on the page.

\begin{figure*}
	\centering
	\includegraphics[width = 0.95\textwidth]{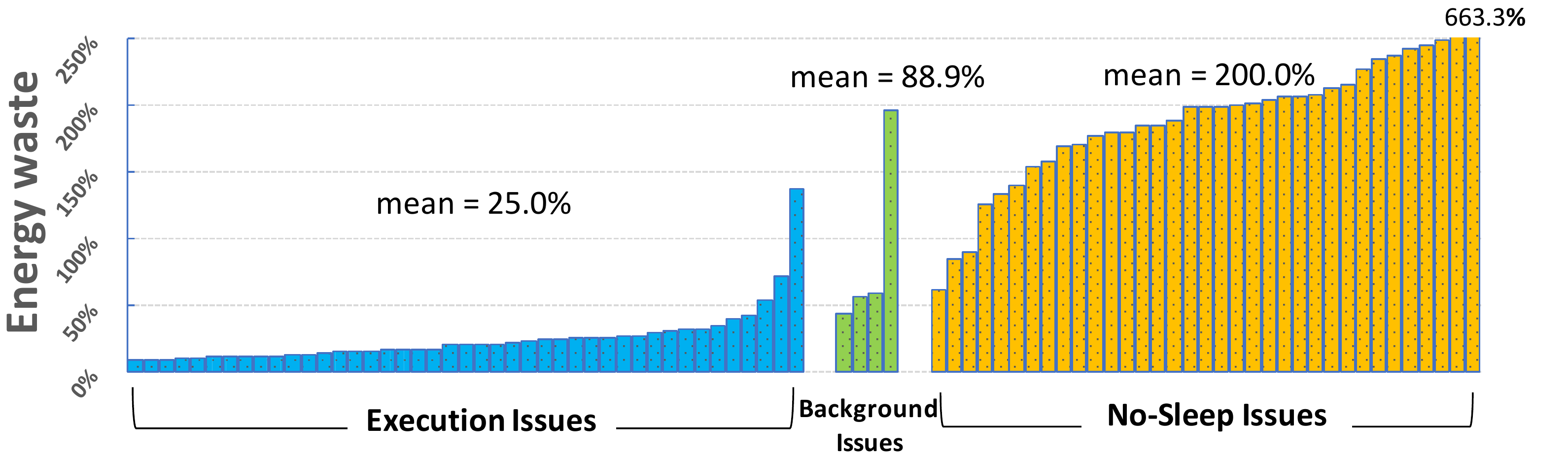}
	\caption{\textbf{The energy waste of detected issues of different types}} \label{figure:energy_waste}
\end{figure*}

 We only have four ($4/83=4.8\%$) background issues, which indicates that operating system is competent in clearing the influence of backgrounded apps at \texttt{BACKGROUND} stage. However, backgrounded apps may still suffer from no-sleep issues: \texttt{Normal}, \texttt{Network Fail} and \texttt{Flight Mode} provoke 6, 17, 13 no-sleep issues respectively. Of particular note is that even though \texttt{Non-background} and \texttt{Normal} are running under same context, the former inclines to provoke more no-sleep issues. In our test, it brings on 17 issues.
 This result implies that special contexts (and \texttt{Non-background}) tend to incur anomalous behaviours of apps, such as bad use of wake lock, and thus cause more no-sleep issues.

 Figure \ref{figure:distribution_detected_energy_issues} also presents that 37.3\% (31 out of 83) energy issues can only be triggered under \texttt{Network Fail} and \texttt{Flight Mode}. This confirms the Finding 2: special context, such as network fail, hides a significant number of serious energy issues. 
 

   \begin{figure}
   	\vspace{-4mm}
   	\centering
   	\includegraphics[width = 0.35\textwidth]{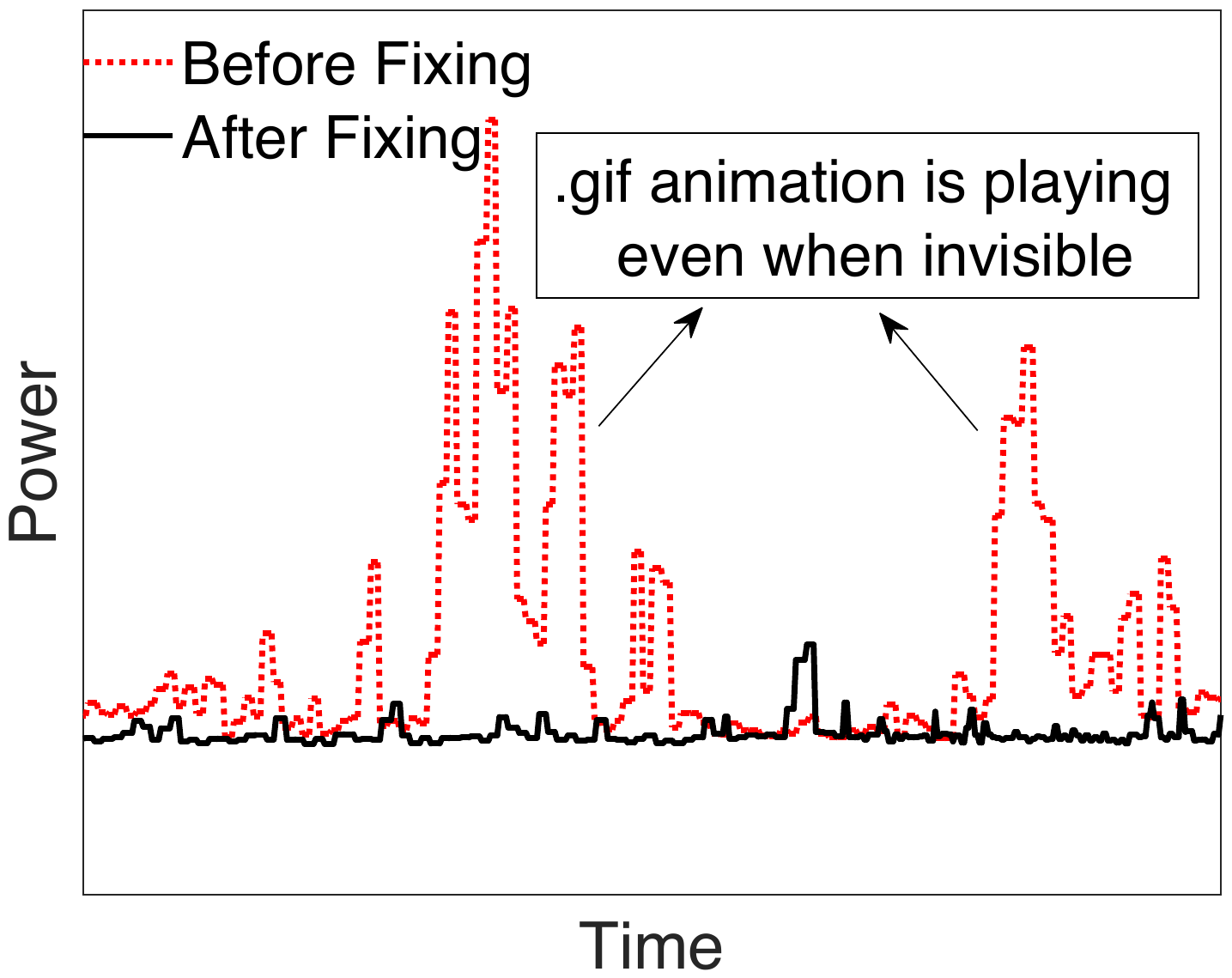}
   	\caption{Power traces before and after fixing the energy issues in Leisure.} \label{fig:leisure_powertrace}
   	\vspace{-4mm}
   \end{figure}

 \subsection{Energy waste of the detected issues}\label{section_energy_waste}

 Figure \ref{figure:energy_waste} shows the energy waste of detected energy issues. The energy waste is calculated using Equation (\ref{formulas_energywaste}). 

 \begin{equation}
 \label{formulas_energywaste}
 w =( \frac{e_x} {e_n} -1) \times 100\%
 \end{equation}
 
 $w$ is the energy waste of the issue, $e_x$ is average power of the corresponding stage in the test case with the corresponding issue (\texttt{EXECUTION} stage for execution issues, \texttt{BACKGROUND} stage for background issues, \texttt{SCREEN-OFF} stage for no-sleep issues). $e_n$ is ``normal" energy cost. We define ``normal" energy cost individually for different issues. For background and no-sleep issues, we adopt average power at \texttt{IDLE} and \texttt{PRE-OFF} stage as normal cost, respectively. For execution issues, we use mean value of average powers of \texttt{EXECUTION} stage in test cases in the same app category, as normal cost.

  \begin{figure}
  		\vspace{-4mm}
  	\centering
  	\includegraphics[width = 0.35\textwidth]{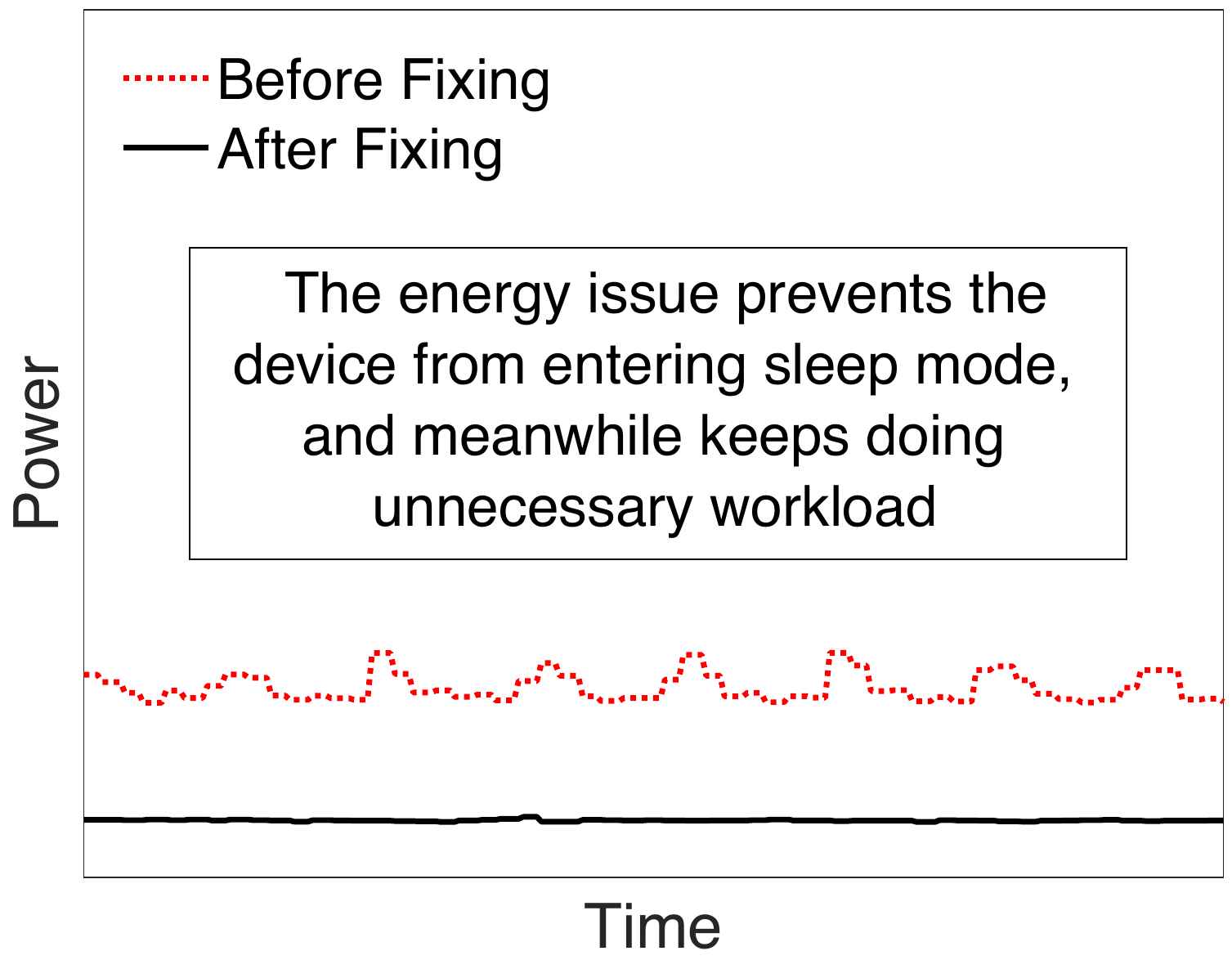}
  	\caption{Power traces before and after fixing the energy issues in Vanilla.} \label{fig:vanilla_powertrace}
  	\vspace{-4mm}
  \end{figure}
 
 The experimental result shows that, the energy waste of execution (Exe) issues is 25.0\% on average and up to 137.6\%, the energy waste of background (BG) issues is 88.9\% on average and at maximum 196.2\%, the values for no-sleep (NS) issues are 200.0\% and 663.3\%, separately. Overall, the average energy waste of the issues is 101.7\%.
 
\subsection{How can our framework benefit developers?}
 After identifying the energy issue, our framework generates a test report to help fix the issue. The report includes:

\vspace{-1mm}
 \begin{itemize}[itemsep = 4pt, topsep = 4pt]
     \item The information on input sequence and running context, and screen-casting video recording the test case. Having them, developers can analyse the manifestation and symptoms of the issue.
     \vspace{-2mm}
     
     \item The visualized power trace of the test case. With it,  developers will have an intuitive view on power consumption of the issue. 
     \vspace{-2mm}
     
     \item The method-level execution trace provided by Android Trace View. The issue reports in empirical study showed that this information can always ($100\%$) help find the faulty code.
     \vspace{-2mm}
     
     \item The rationale behind our testing framework. Developers will understand why and how the testing framework works. 
     \vspace{-2mm}
\end{itemize}

We also conducted a case study, showing how to utilize the test report to diagnose and fix energy issues. We take the execution issue in Leisure and the no-sleep issue in Vanilla (as we listed in Table \ref{table:energy_issue_examples}) as two examples.  We reproduce the issues, observe the symptoms, check the power traces (the red dash lines in Figure \ref{fig:leisure_powertrace} and \ref{fig:vanilla_powertrace}). We then find the frequently-called methods in the source code that result in energy waste, and refactor the code to remove the unnecessary workload and release the wake-lock after use. As we can see, the power traces (black full lines in Figure \ref{fig:leisure_powertrace} and \ref{fig:vanilla_powertrace}) after fixing are much flatter or lower than those before fixing.





\section{Related Work}

Our work is related to multiple lines of research work. We will discuss in three areas: \textit{understanding energy issues}; \textit{detecting \& diagnosing energy issues}; \textit{fixing energy issues \& optimizing software}.  

\paragraph{Understanding Energy Issues}


The first empirical study on characteristics of energy issues in the system of mobile device was done by Pathak et al. \cite{Pathak_empiricalStudy_energyIssue}. They mined over 39,000 posts from four online mobile user forums and mobile OS bug repositories, and studied the categorization and manifestation of energy issues. These studied issues involve multiple layers across the system stack of mobile devices, from hardware, operating systems to applications. In comparison, our empirical study in this paper focuses on app-level energy issues.

The above style of empirical study that mines the data in project repositories has been widely applied. For example, to characterize performance issues, a large body of research has been done for PC and server side software \cite{Jin_empirical_perf1,Nistor_empirical_perf2,Zaman_empirical_perf3,Liu_empirical_perf}. The most related study is conducted by Liu et al. \cite{Liu_empirical_perf} that investigates performance issues (including energy issues) in Android applications. However, the number of the studied energy issues is very small compared with our study. More importantly, our study is only targeted at energy issues, so our findings are more specialized and comprehensive on energy issues. 


\paragraph{Detecting \& Diagnosing Energy Issues}


To detect abnormal energy use of mobile apps, researchers use operating system and hardware features as predictors to infer energy information at device, component, virtual machine or application level  \cite{infocom_paper,Kansal_powerofvm,Pathak_whereisenergy,ubicomp_app_profiling,network_monitoring,AppScope,PowerForecaster}.


 Shuai et al. \cite{HaoShuai:2013:EstMobileApp} and Ding et al. \cite{sourceline_energy} proposed approaches to obtaining energy information at source line level. The former requires the specific energy profile of the target systems, which may not be accessible to developers. The latter utilizes advanced measurement techniques to obtain source line energy cost. 


The work \cite{Banerjee_energy_bug_detection} is the most relevant to ours. Even though their technology is competent in detecting background and no-sleep issues, two key differences prevent it from uncovering most serious energy issues detected in our test. The first difference is that their work could not deal with execution issues properly, as we discussed in Section \ref{section_introduction}. The second difference is that their work did not take the special running context into consideration. As shown in our experiment, 37.3\% (31 out of 83) energy issues can only manifest themselves under special context. In summary, owing to our high-quality empirical study and well-designed testing technology, our framework surpasses the state-of-the-art in detecting all kinds of energy issues.

\paragraph{Fixing Energy Issues \& Optimizing Software}

A large amount of research effort on energy-saving for mobile devices has been focused on the main hardware components, such as the CPU, display and network interface. The CPU-related techniques involve dynamic voltage and frequency scaling~\cite{DVFS2}, heterogeneous architecture \cite{Reflex_Lin,GreeDroid_Goulding,Li_big.Little} and computation offloading~\cite{computation_offloading,nice_friend}. Techniques targeting the display include dynamic frame rate tuning~\cite{dynamic_framerate_tunning}, dynamic resolution tuning \cite{Resolution_Scaling} and tone-mapping based back-light scaling~\cite{tone_mapping,tone_mapping2}.  Network-related techniques try to exploit idle and deep sleep opportunities~\cite{network1,network2}, shape the traffic patterns~\cite{network_3,network4}, trade-off energy against other criteria~\cite{Rate-distortion-energy_Tradeoff,Energy-delay_tradeoff,Energy_Performance_Trade_Off} and so on. Such work attempts to reduce energy dissipation by optimizing the hardware usage; on the other hand, several pieces of work aim at designing new hardware and devices \cite{hardwaredisign1,hardwaredesign2}.





Two pieces of work \cite{GeneticImprovement,Seeds} provide systematic approaches to optimizing software source code. In the former, Boddy et al. attempted to decrease the energy consumption of software by handling code as if it were genetic material so as to evolve to be more energy-efficient. In the latter, Irene et al. proposed a framework to optimize Java applications by iteratively searching for more energy-saving implementations in the design space.

\section{Conclusion}\label{section_conclusion}
In this paper, we conducted an empirical study on software energy issues in 27 well-maintained open-source mobile apps. Our study revealed root causes and manifestation of energy issues.
Inspired by this study, we fully implemented a novel testing framework for detecting energy issues. It first statically analyses the source code of app subjects and then extracts the candidate input-sequences with large probability of causing energy issues. We also devised several artificial runtime contexts that can expose deeply-hidden energy issues. Our framework effectively examines apps with the inputs and contexts under a systematic scheme, and then automatically identifies energy issues from power traces.
A large-scale experimental evaluation showed that our framework is capable of detecting a large number of energy issues, most of which existing techniques cannot handle. These issues on average double the energy cost of the apps. Finally, we showed how  developers can utilize our test reports to fix the issues.
\\

\bibliographystyle{abbrv}
{\scriptsize
\bibliography{sigproc} }

\begin{thebibliography}{10}

\bibitem{tone_mapping2}
B.~Anand, K.~Thirugnanam, J.~Sebastian, P.~G. Kannan, A.~L. Ananda, M.~C. Chan,
  and R.~K. Balan.
\newblock Adaptive display power management for mobile games.
\newblock In {\em Proceedings of the 9th International Conference on Mobile
  Systems, Applications, and Services}, MobiSys '11, pages 57--70, New York,
  NY, USA, 2011. ACM.

\bibitem{Banerjee_energy_bug_detection}
A.~Banerjee, L.~K. Chong, S.~Chattopadhyay, and A.~Roychoudhury.
\newblock Detecting energy bugs and hotspots in mobile apps.
\newblock In {\em Proceedings of the 22Nd ACM SIGSOFT International Symposium
  on Foundations of Software Engineering}, FSE 2014, pages 588--598, New York,
  NY, USA, 2014. ACM.

\bibitem{GeneticImprovement}
B.~R. Bruce, J.~Petke, and M.~Harman.
\newblock Reducing energy consumption using genetic improvement.
\newblock In {\em Proceedings of the 2015 Annual Conference on Genetic and
  Evolutionary Computation}, GECCO '15, pages 1327--1334, New York, NY, USA,
  2015. ACM.

\bibitem{Energy_Performance_Trade_Off}
D.~H. Bui, Y.~Liu, H.~Kim, I.~Shin, and F.~Zhao.
\newblock Rethinking energy-performance trade-off in mobile web page loading.
\newblock In {\em Proceedings of the 21st Annual International Conference on
  Mobile Computing and Networking}, MobiCom '15, pages 14--26, New York, NY,
  USA, 2015. ACM.

\bibitem{network_3}
C.~Chiasserini and R.~Rao.
\newblock Improving battery performance by using traffic shaping techniques.
\newblock {\em Selected Areas in Communications, IEEE Journal on},
  19(7):1385--1394, Jul 2001.

\bibitem{DVFS2}
V.~Devadas and H.~Aydin.
\newblock On the interplay of voltage/frequency scaling and device power
  management for frame-based real-time embedded applications.
\newblock {\em IEEE Transactions on Computers}, 61(1):31--44, Jan 2012.

\bibitem{DbScan}
M.~Ester, H.-P. Kriegel, J.~Sander, and X.~Xu.
\newblock A density-based algorithm for discovering clusters a density-based
  algorithm for discovering clusters in large spatial databases with noise.
\newblock In {\em Proceedings of the Second International Conference on
  Knowledge Discovery and Data Mining}, KDD'96, pages 226--231. AAAI Press,
  1996.

\bibitem{GreeDroid_Goulding}
N.~Goulding-Hotta, J.~Sampson, G.~Venkatesh, S.~Garcia, J.~Auricchio, P.~Huang,
  M.~Arora, S.~Nath, V.~Bhatt, J.~Babb, S.~Swanson, and M.~Taylor.
\newblock The {GreenDroid} mobile application processor: An architecture for
  silicon's dark future.
\newblock {\em Micro, IEEE}, 31(2):86--95, March 2011.

\bibitem{HaoShuai:2013:EstMobileApp}
S.~Hao, D.~Li, W.~G.~J. Halfond, and R.~Govindan.
\newblock Estimating mobile application energy consumption using program
  analysis.
\newblock In {\em Proceedings of the 2013 International Conference on Software
  Engineering}, ICSE '13, pages 92--101, Piscataway, NJ, USA, 2013. IEEE Press.

\bibitem{Resolution_Scaling}
S.~He, Y.~Liu, and H.~Zhou.
\newblock Optimizing smartphone power consumption through dynamic resolution
  scaling.
\newblock In {\em Proceedings of the 21st Annual International Conference on
  Mobile Computing and Networking}, MobiCom '15, pages 27--39, New York, NY,
  USA, 2015. ACM.

\bibitem{dynamic_framerate_tunning}
C.~Hwang, S.~Pushp, C.~Koh, J.~Yoon, Y.~Liu, S.~Choi, and J.~Song.
\newblock Raven: Perception-aware optimization of power consumption for mobile
  games.
\newblock In {\em Proceedings of the 23rd Annual International Conference on
  Mobile Computing and Networking}, MobiCom '17, pages 422--434, New York, NY,
  USA, 2017. ACM.

\bibitem{tone_mapping}
A.~Iranli and M.~Pedram.
\newblock {DTM}: Dynamic tone mapping for backlight scaling.
\newblock In {\em Proceedings of the 42Nd Annual Design Automation Conference},
  DAC '05, pages 612--617, New York, NY, USA, 2005. ACM.

\bibitem{Jin_empirical_perf1}
G.~Jin, L.~Song, X.~Shi, J.~Scherpelz, and S.~Lu.
\newblock Understanding and detecting real-world performance bugs.
\newblock In {\em Proceedings of the 33rd ACM SIGPLAN Conference on Programming
  Language Design and Implementation}, PLDI '12, pages 77--88, New York, NY,
  USA, 2012. ACM.

\bibitem{Kansal_powerofvm}
A.~Kansal, F.~Zhao, J.~Liu, N.~Kothari, and A.~A. Bhattacharya.
\newblock Virtual machine power metering and provisioning.
\newblock In {\em Proceedings of the 1st ACM Symposium on Cloud Computing},
  SoCC '10, pages 39--50, New York, NY, USA, 2010. ACM.

\bibitem{infocom_paper}
J.~Koo, K.~Lee, W.~Lee, Y.~Park, and S.~Choi.
\newblock Batttracker: Enabling energy awareness for smartphone using li-ion
  battery characteristics.
\newblock In {\em IEEE INFOCOM 2016 - The 35th Annual IEEE International
  Conference on Computer Communications}, pages 1--9, April 2016.

\bibitem{computation_offloading}
K.~Kumar and Y.~Lu.
\newblock Cloud computing for mobile users: Can offloading computation save
  energy?
\newblock {\em Computer}, 43(4):51--56, April 2010.

\bibitem{ubicomp_app_profiling}
S.~Lee, C.~Yoon, and H.~Cha.
\newblock User interaction-based profiling system for android application
  tuning.
\newblock In {\em Proceedings of the 2014 ACM International Joint Conference on
  Pervasive and Ubiquitous Computing}, UbiComp '14, pages 289--299, New York,
  NY, USA, 2014. ACM.

\bibitem{source_level_practice}
D.~Li and W.~G.~J. Halfond.
\newblock An investigation into energy-saving programming practices for android
  smartphone app development.
\newblock In {\em Proceedings of the 3rd International Workshop on Green and
  Sustainable Software}, GREENS 2014, pages 46--53, New York, NY, USA, 2014.
  ACM.

\bibitem{sourceline_energy}
D.~Li, S.~Hao, W.~G.~J. Halfond, and R.~Govindan.
\newblock Calculating source line level energy information for {A}ndroid
  applications.
\newblock In {\em Proceedings of the 2013 International Symposium on Software
  Testing and Analysis}, ISSTA 2013, pages 78--89, New York, NY, USA, 2013.
  ACM.

\bibitem{Li_big.Little}
X.~Li, G.~Chen, and W.~Wen.
\newblock Energy-efficient execution for repetitive app usages on big.little
  architectures.
\newblock In {\em Proceedings of the 54th Annual Design Automation Conference
  2017}, DAC '17, pages 44:1--44:6, New York, NY, USA, 2017. ACM.

\bibitem{source_level}
X.~Li and J.~P. Gallagher.
\newblock A source-level energy optimization framework for mobile applications.
\newblock In {\em 2016 IEEE 16th International Working Conference on Source
  Code Analysis and Manipulation (SCAM)}, pages 31--40, Oct 2016.

\bibitem{Reflex_Lin}
F.~X. Lin, Z.~Wang, R.~LiKamWa, and L.~Zhong.
\newblock Reflex: Using low-power processors in smartphones without knowing
  them.
\newblock {\em SIGPLAN Not.}, 47(4):13--24, Mar. 2012.

\bibitem{network1}
J.~Liu and L.~Zhong.
\newblock Micro power management of active 802.11 interfaces.
\newblock In {\em Proceedings of the 6th International Conference on Mobile
  Systems, Applications, and Services}, MobiSys '08, pages 146--159, New York,
  NY, USA, 2008. ACM.

\bibitem{Liu_wakelock}
Y.~Liu, C.~Xu, S.~Cheung, and V.~Terragni.
\newblock Understanding and detecting wake lock misuses for android
  applications.
\newblock In {\em Proceedings of the 24th ACM SIGSOFT International Symposium
  on the Foundations of Software Engineering, {FSE} 2016}, 2016.

\bibitem{Liu_empirical_perf}
Y.~Liu, C.~Xu, and S.-C. Cheung.
\newblock Characterizing and detecting performance bugs for smartphone
  applications.
\newblock In {\em Proceedings of the 36th International Conference on Software
  Engineering}, ICSE 2014, pages 1013--1024, New York, NY, USA, 2014. ACM.

\bibitem{Seeds}
I.~Manotas, L.~Pollock, and J.~Clause.
\newblock Seeds: A software engineer's energy-optimization decision support
  framework.
\newblock In {\em Proceedings of the 36th International Conference on Software
  Engineering}, ICSE 2014, pages 503--514, New York, NY, USA, 2014. ACM.

\bibitem{Event_flow_graph}
A.~Memon, I.~Banerjee, and A.~Nagarajan.
\newblock Gui ripping: reverse engineering of graphical user interfaces for
  testing.
\newblock In {\em 10th Working Conference on Reverse Engineering, 2003. WCRE
  2003. Proceedings.}, pages 260--269, Nov 2003.

\bibitem{PowerForecaster}
C.~Min, Y.~Lee, C.~Yoo, S.~Kang, I.~Hwang, and J.~Song.
\newblock Powerforecaster: Predicting power impact of mobile sensing
  applications at pre-installation time.
\newblock {\em GetMobile: Mobile Comp. and Comm.}, 20(1):30--33, July 2016.

\bibitem{Nistor_empirical_perf2}
A.~Nistor, T.~Jiang, and L.~Tan.
\newblock Discovering, reporting, and fixing performance bugs.
\newblock In {\em Proceedings of the 10th Working Conference on Mining Software
  Repositories}, MSR '13, pages 237--246, Piscataway, NJ, USA, 2013. IEEE
  Press.

\bibitem{Pathak_empiricalStudy_energyIssue}
A.~Pathak, Y.~C. Hu, and M.~Zhang.
\newblock Bootstrapping energy debugging on smartphones: A first look at energy
  bugs in mobile devices.
\newblock In {\em Proceedings of the 10th ACM Workshop on Hot Topics in
  Networks}, HotNets-X, pages 5:1--5:6, New York, NY, USA, 2011. ACM.

\bibitem{Pathak_whereisenergy}
A.~Pathak, Y.~C. Hu, and M.~Zhang.
\newblock Where is the energy spent inside my app?: Fine grained energy
  accounting on smartphones with {Eprof}.
\newblock In {\em Proceedings of the 7th ACM European Conference on Computer
  Systems}, EuroSys '12, pages 29--42, New York, NY, USA, 2012. ACM.

\bibitem{paper1_use_odroid}
A.~{Pathania}, Q.~{Jiao}, A.~{Prakash}, and T.~{Mitra}.
\newblock Integrated cpu-gpu power management for 3d mobile games.
\newblock In {\em Proceedings of the 51st Annual Design Automation Conference
  on}, pages 1--6, 2014.

\bibitem{network4}
C.~Poellabauer and K.~Schwan.
\newblock Energy-aware traffic shaping for wireless real-time applications.
\newblock In {\em Real-Time and Embedded Technology and Applications Symposium,
  2004. Proceedings. RTAS 2004. 10th IEEE}, pages 48--55, May 2004.

\bibitem{network_monitoring}
L.~Ravindranath, J.~Padhye, S.~Agarwal, R.~Mahajan, I.~Obermiller, and
  S.~Shayandeh.
\newblock Appinsight: Mobile app performance monitoring in the wild.
\newblock In {\em Proceedings of the 10th USENIX Conference on Operating
  Systems Design and Implementation}, OSDI'12, pages 107--120, Berkeley, CA,
  USA, 2012. USENIX Association.

\bibitem{Energy-delay_tradeoff}
A.~Sehati and M.~Ghaderi.
\newblock Energy-delay tradeoff for request bundling on smartphones.
\newblock In {\em IEEE INFOCOM 2017 - IEEE Conference on Computer
  Communications}, pages 1--9, May 2017.

\bibitem{network2}
J.~Sorber, N.~Banerjee, M.~D. Corner, and S.~Rollins.
\newblock Turducken: Hierarchical power management for mobile devices.
\newblock In {\em Proceedings of the 3rd International Conference on Mobile
  Systems, Applications, and Services}, MobiSys '05, pages 261--274, New York,
  NY, USA, 2005. ACM.

\bibitem{nice_friend}
K.~Sucipto, D.~Chatzopoulos, S.~Kosta±, and P.~Hui.
\newblock Keep your nice friends close, but your rich friends closer —
  computation offloading using nfc.
\newblock In {\em IEEE INFOCOM 2017 - IEEE Conference on Computer
  Communications}, pages 1--9, May 2017.

\bibitem{hardwaredisign1}
T.~Tuan, S.~Kao, A.~Rahman, S.~Das, and S.~Trimberger.
\newblock A 90nm low-power {FPGA} for battery-powered applications.
\newblock In {\em Proceedings of the 2006 ACM/SIGDA 14th International
  Symposium on Field Programmable Gate Arrays}, FPGA '06, pages 3--11, New
  York, NY, USA, 2006. ACM.

\bibitem{hardwaredesign2}
L.~Wang, M.~French, A.~Davoodi, and D.~Agarwal.
\newblock {FPGA} dynamic power minimization through placement and routing
  constraints.
\newblock {\em EURASIP J. Embedded Syst.}, 2006(1):7--7, Jan. 2006.

\bibitem{paper2_use_odroid}
H.~Yamamoto, T.~Hirano, K.~Muto, H.~Mikami, T.~Goto, D.~Hillenbrand,
  M.~Takamura, K.~Kimura, and H.~Kasahara.
\newblock Oscar compiler controlled multicore power reduction on android
  platform.
\newblock In {\em Languages and Compilers for Parallel Computing}, pages
  155--168, Cham, 2014. Springer International Publishing.

\bibitem{Rate-distortion-energy_Tradeoff}
Z.~Yan and C.~W. Chen.
\newblock Rnb: Rate and brightness adaptation for rate-distortion-energy
  tradeoff in http adaptive streaming over mobile devices.
\newblock In {\em Proceedings of the 22Nd Annual International Conference on
  Mobile Computing and Networking}, MobiCom '16, pages 308--319, New York, NY,
  USA, 2016. ACM.

\bibitem{AppScope}
C.~Yoon, D.~Kim, W.~Jung, C.~Kang, and H.~Cha.
\newblock Appscope: Application energy metering framework for android
  smartphones using kernel activity monitoring.
\newblock In {\em Proceedings of the 2012 USENIX Conference on Annual Technical
  Conference}, USENIX ATC'12, pages 36--36, Berkeley, CA, USA, 2012. USENIX
  Association.

\bibitem{Zaman_empirical_perf3}
S.~Zaman, B.~Adams, and A.~E. Hassan.
\newblock A qualitative study on performance bugs.
\newblock In {\em Proceedings of the 9th IEEE Working Conference on Mining
  Software Repositories}, MSR '12, pages 199--208, Piscataway, NJ, USA, 2012.
  IEEE Press.

\end{thebibliography}
\end{document}